\documentclass[aps,prd,preprint,floats,epsf,superscriptaddress,nofootinbib]{revtex4}
\usepackage{filecontents}
\pdfoutput=1
\usepackage{textcomp}
\usepackage{amssymb}
\usepackage{anysize}
\usepackage{bold-extra}
\usepackage{enumerate}
\usepackage{graphicx}
\usepackage{lscape}
\usepackage{mathrsfs}
\usepackage{multirow}
\usepackage{hhline}
\usepackage{makecell}
\usepackage[final]{pdfpages}
\usepackage{rotating}
\usepackage{subfig}
\usepackage{url}
\usepackage{wrapfig}
\usepackage{amsmath}
\usepackage{color}
\usepackage{xcolor}

\usepackage{fancyhdr}
\usepackage{verbatim}
\usepackage{epstopdf}
\usepackage[section]{placeins}
\usepackage[normalem]{ulem} % this one is for cross out text
\usepackage[colorlinks=true,
            linkcolor=red,
            urlcolor=blue,
            citecolor=blue]{hyperref}
\def\nn{\nonumber}
\def\bea{\begin{eqnarray}}
\def\eea{\end{eqnarray}}
\def\ba{\begin{eqnarray}}
\def\ea{\end{eqnarray}}
\def\be{\begin{equation}}
\def\ee{\end{equation}}
\def\beq{\begin{equation}}
\def\eeq{\end{equation}}

\newcommand{\tf}{\texorpdfstring}

\newcommand{\gev}{~\text{GeV}}
\newcommand{\tev}{~\text{TeV}}

\newcommand{\abi}{~\text{ab}^{-1}}

\begin{document}

\baselineskip=17pt

\title{Searching for a charged Higgs boson with both \\
$H^{\pm}W^{\mp}Z$ and $H^{\pm}tb$ couplings at the LHC}

\author{Jian-Yong Cen}
\email[e-mail:]{cenjy@sxnu.edu.cn}
\affiliation{School of Physics and Information Engineering, Shanxi Normal University, Linfen, Shanxi 041004}

\author{Jung-Hsin Chen}
\email[e-mail:]{lewis02030405@gmail.com}
\affiliation{Department of Physics, National Taiwan University, Taipei, Taiwan 10617}

\author{Xiao-Gang He}
\email[e-mail: ]{hexg@phys.ntu.edu.tw}
\affiliation{Tsung-Dao Lee Institute $\&$ SKLPPC,  School of Physics and Astronomy, Shanghai Jiao Tong University,
800 Dongchuan Rd.,  Shanghai 200240}
\affiliation{Department of Physics, National Taiwan University, Taipei, Taiwan 10617}
\affiliation{Physics Division, National Center for Theoretical Sciences, Hsinchu, Taiwan 30013}

\author{Gang Li}
\email[e-mail: ]{gangli@phys.ntu.edu.tw}
\affiliation{Department of Physics, National Taiwan University, Taipei, Taiwan 10617}

\author{Jhih-Ying Su}
\email[e-mail:]{b02202013@ntu.edu.tw }
\affiliation{Department of Physics, National Taiwan University, Taipei, Taiwan 10617}

\author{Wei Wang}
\email[e-mail: ]{wei.wang@sjtu.edu.cn}
\affiliation{INPAC, SKLPPC, School of Physics and Astronomy,  Shanghai Jiao Tong University,
800 Dongchuan Rd., Shanghai 200240}

\date{\today}

\vskip 1cm
\begin{abstract}
In certain new physics scenarios, a singly charged Higgs boson can couple to both fermions and $W^\pm Z$ at tree level. We develop new strategies beyond current experimental  searches using $pp\to jjH^\pm$, $H^\pm \to tb $ at the Large Hadron Collider (LHC). With the effective $H^\pm W^\mp Z$ and $H^\pm tb$ couplings we perform a model-independent analysis at the collision energy $\sqrt{s}=13$~TeV with the integrated luminosity of $3~\text{ab}^{-1}$. We derive the discovery prospects and exclusion limits for the charged Higgs boson in the mass range from 200~GeV to 1~TeV. With $|F_{WZ}|,|A_t|\sim 0.5-1.0$ and $300~\text{GeV}\lesssim m_{H^\pm}\lesssim 400~\text{GeV}$, we point out that a discovery significance of $5\sigma$ can be achieved. The constraints and projected sensitivities are also discussed in a realistic model, i.e., the modified Georgi-Machacek model without custodial symmetry. Our proposed search would provide direct evidence for a charged Higgs boson $H^\pm$ that couples to $W^\pm Z$ and $tb$, and has better sensitivity to the couplings of  $H^\pm W^\mp Z$ and $H^\pm tb$ than current searches.

\end{abstract}

%\pacs{PACS numbers: }

\maketitle
    
%%%%%%%

%\newpage

\section{Introduction}

The discovery of the Higgs boson with a mass around 125~GeV~\cite{Aad:2012tfa,Chatrchyan:2012xdj} at the  Large Hadron Collider (LHC) has provided much information  on  electroweak symmetry breaking (EWSB) and  particle mass generation. All available measurements of the couplings of the Higgs boson~\cite{Khachatryan:2016vau,ATLAS:2018doi,CMS:2018lkl} are consistent with standard model (SM) with a minimal scalar sector of one Higgs doublet $H$. In the SM, $H$ transforms under the  electroweak gauge group $SU(2)_L\times U(1)_Y$ as $(2, -1/2)$~\footnote{We use the notation $Q=T_3+Y$.} and can be written as $H = ((v_H + h + i I_H)/\sqrt{2}, h^-)^T$. The non-zero vacuum expectation value (VEV) $v_H$ is the source to induce  the EWSB and generate the mass.  $h^\pm$ and $I_H$ are ``eaten'' by $W^\pm$ and $Z$ boson as their longitudinal components and therefore only $h$ is a physical Higgs boson. Is the Higgs boson  discovered by the LHC really the one in the SM or does it have new interactions? Are there additional Higgs bosons? These are important questions in particle physics today.

In models extended with additional Higgs multiplets,  one can achieve a richer particle spectrum in the scalar sector. By adding a $SU(2)_L$ scalar singlet with hypercharge equal to zero, one can obtain another neutral Higgs boson that  may mix with $h$. While for a charged Higgs boson the situation is different. Scalar singlet(s) with non-trivial hypercharge(s)~\footnote{Model-independent studies of singlet scalar with non-trivial hypercharge can be found in Refs.~\cite{Cao:2017ffm,Cao:2018ywk}.} or higher multiplet(s) beyond the SM are needed. In the two-Higgs-doublet models (2HDMs)~\cite{Branco:2011iw}, there is one physical charged Higgs boson $H_D^\pm$ (the subscript ``D'' denotes doublet) in the particle spectrum. The couplings of $H_D^\pm$ to the fermions are proportional to fermion masses.  It was noticed~\cite{Grifols:1980uq,Gunion:1989we} that in models with only Higgs doublets, there  is no tree-level coupling of charged Higgs boson to $W^\pm$ and $Z$ bosons. Although such interaction can be generated at loop level, it is usually suppressed~\cite{Kanemura:1999tg,Moretti:2015tva,Abbas:2018pfp}. Phenomenology of charged Higgs boson in the 2HDMs has been discussed thoroughly, see Ref.~\cite{Akeroyd:2016ymd} for a recent review.

If   SM is extended with higher $SU(2)_L$ representation of scalar field(s), the  charged Higgs boson couples to $W^\pm Z$  at   tree level. Models with $SU(2)_L$ scalar triplet(s) are attractive since they can generate  the neutrino mass~\cite{Lazarides:1980nt,Magg:1980ut,Mohapatra:1980yp,Cheng:1980qt} and provide dark matter candidate~\cite{Cirelli:2005uq}, etc. Nevertheless, in the models with one complex/real triplet, the $\rho=m_W^2/m_Z^2\cos^2\theta_W$ parameter deviates from unity, which  constrains   the triplet VEV stringently. If  one complex and one real triplets are introduced  and  the triplets have the same VEVs (In our convention, it corresponds to the case that the real triplet VEV is $1/\sqrt{2}$ of the complex triplet VEV, see more details in Section~\ref{sec:model}), the relation $\rho=1$ can be maintained at   tree level~\cite{Georgi:1985nv}. This is protected by $SU(2)_L\times SU(2)_R\to SU(2)_C$ of the Higgs potential surviving after the EWSB~\cite{Chanowitz:1985ug} similar with  the SM. Here $SU(2)_C$ represents the custodial symmetry~\cite{Sikivie:1980hm}. An example of this type is  Georgi-Machacek (GM) model, in which the  triplet VEV can be large. There are two singly charged Higgs bosons $H_3^\pm$ and $H_5^\pm$ in the GM model, which transform as 3-plet and 5-plet under $SU(2)_C$, respectively~\cite{Chanowitz:1985ug}. Due to the custodial symmetry, $H_3^\pm$ and $H_5^\pm$ couple to fermions and $W^\mp Z$ separately and there is no mixing between them. Besides, both the couplings of $H_3^\pm$ to fermions and $H_5^\pm$ to $W^\mp Z$ are proportional to the triplet VEV~\cite{Chiang:2012cn,Hartling:2014zca}.

Experimental searches for singly charged Higgs bosons have been carried out at high-energy colliders. The search strategies depend on the couplings of   charged Higgs boson to SM particles. The LEP experiments have excluded the charged Higgs boson in the type-II 2HDM with mass below 80~GeV~\cite{Abbiendi:2013hk}. At the LHC, the  $H_D^\pm$ in the 2HDMs is primarily produced from top quark decay when lighter than top quark, i.e. $m_{H_D^+}<m_t$ or in the  association production  with top quark if $m_{H_D^+}>m_t$. Various decay channels including~\footnote{For simplicity, the notation ``$f_1f_2$'' denotes $f_1\bar{f}_2$ and $\bar{f}_1f_2$, where $f_1f_2=tb,\tau\nu,cs,cb$.} $tb$~\cite{Aaboud:2018cwk,Aad:2015typ,Khachatryan:2015qxa} $\tau\nu$~\cite{Aaboud:2018gjj,CMS:2016szv,Aad:2014kga,Khachatryan:2015qxa}, $cs$~\cite{Khachatryan:2015uua} and $cb$~\cite{Sirunyan:2018dvm} have been explored for $H_D^\pm$ in the mass range from 90~GeV to 2~TeV with no significant excess observed. For the   $H_5^\pm $ in the GM model, it has been sought in the vector-boson-fusion (VBF) process $pp\to jjH_5^\pm$, $H_5^\pm\to W^\pm Z$~\cite{Sirunyan:2017sbn,Aad:2015nfa} at the LHC for $200\gev\leq m_{H_5^\pm}\leq1\tev$.

However, an interesting  scenario in which $H^\pm$ couples to both fermions and $W^\mp Z$ significantly  has not been extensively explored. Such a charged Higgs boson does not exist in many   popular Higgs extended models, such as 2HDMs and GM model. Nevertheless,  once  a singly charged Higgs boson was discovered in one   existing channel,    its couplings to   all SM particles are requested  to verify its nature. 
In this scenario, new production and decay channels can be utilized.
Since the couplings of $H^\pm$  to fermions are usually proportional to the fermion masses, its couplings to the third generation quarks are more relevant. The property of charged Higgs boson with both couplings to fermions and $W^\mp Z$ can be revealed through the processes: (1) $pp\to jjH^\pm$, $H^\pm\to tb$; (2) $pp\to tH^\pm$, $H^\pm\to W^\pm Z$. In this work, we will study the first process $pp\to jjH^\pm$, $H^\pm\to tb$ by investigating the sensitivity to the effective couplings of $H^\pm $ to $W^\mp Z$ and $tb$ and discussing the implications in a realistic modified GM model~\cite{Cen:2018wye,Blasi:2017xmc}~\footnote{For studies on neutral Higgs couplings to gauge bosons in the modified GM model, see Refs.~\cite{Chiang:2018fqf,Chiang:2018irv}.} Whereas, we emphasize that our results are model-independent and can be applied to other models with such charged Higgs boson(s). The second process is left for a future work

This paper is arranged as follows. In Section~\ref{sec:effective}, the effective $H^\pm W^\mp Z$ and $H^\pm tb$ couplings and   decay branching ratios are discussed.  
With the  effective operators, 
we  perform a model-independent collider study  with the currently available  constraints  and the projected sensitivity in our proposed search $pp\to jjH^\pm$, $H^\pm\to tb$. In Section~\ref{sec:model}, we discuss the  search of a charged Higgs in the modified GM model and the implications on model parameters. Section~\ref{sec:summary} summarizes our results.

\section{\tf{$H^{\pm}W^{\mp}Z$}{HWZ} and \tf{$H^\pm tb$}{Htb} couplings and Collider Search Strategies}
\label{sec:effective}

The most general effective Lagrangian that describes the interactions of $H^\pm$ with $W^\pm Z$ and $tb$ is  parameterized as~\cite{DiazCruz:2001tn,Barger:1989fj} 
\begin{align}
\label{eq:couplings}
\mathcal{L}_{\text{eff}}=gm_{W}F_{WZ}H^+W^-_{\mu}Z^{\mu}
-\sqrt{2}/v H^+\bar{t}(m_t A_t P_L + m_b A_b P_R)b+h.c., 
\end{align}
where $P_{L/R}=(1\mp \gamma_5)/2$, and $v\simeq 246$ GeV. For $m_{H^+}>m_{t}+m_{b}$, the partial widths of $H^+$ into $W^+Z$ and $t\bar{b}$ are given by
\begin{align}
\Gamma(H^+\to W^+Z)&=\dfrac{m_{H^+}}{16\pi}\lambda^{1/2}(1,x_W,x_Z)g^2|F_{WZ}|^2\big[\dfrac{(1-x_W-x_Z)^2}{4x_Z}+2x_W\big],\nn\\
\Gamma(H^+\to t\bar{b})&=\dfrac{3g^2|A_t|^2m_t^2}{32\pi m_{H^+}m_W^2}(m_{H^+}^2-m_t^2-m_b^2)\lambda^{1/2}(1,x_t,x_b).
\end{align}
Here $g$ is $SU(2)_L$ gauge coupling, $\lambda(x,y,z)=(x-y-z)^2-4yz$ and $x_{W,Z,t,b}=m_{W,Z,t,b}^2/m_{H^+}^2$. The  $H^+$ has the total width
\begin{align}
\Gamma_{H^+}=\Gamma(H^+\to W^+Z)+\Gamma(H^+\to t\bar{b})+\Gamma(H^+\to \text{others}),
\end{align}
where $\Gamma(H^+\to \text{others})$ denotes the partial width of $H^+$ into other final states. Since the couplings of $H^+$ to fermions are proportional to the fermion masses,   the decay widths of $H^+$ into other fermions are much smaller. 
Thus it is plausible to assume that the total width of $H^+$ is saturated  by  decays into  $W^+Z$ and $t\bar{b}$ (minimal total width assumption). Decay branching ratios of $H^\pm \to t\bar{b}$ and $H^\pm\to W^\pm Z$ are expressed as
\begin{align}
\label{eq:decay_br}
&\mathcal{B}_{tb}\equiv \mathcal{B}(H^\pm\to tb)=\dfrac{\Gamma(H^+\to t\bar{b})}{\Gamma(H^+\to t\bar{b})+\Gamma(H^+\to W^+Z)},\\
&\mathcal{B}_{WZ}\equiv \mathcal{B}(H^\pm\to W^\pm Z)=\dfrac{\Gamma(H^+\to W^+Z)}{\Gamma(H^+\to t\bar{b})+\Gamma(H^+\to W^+Z)}.
\end{align}

In the following, the effective couplings in Eq.~\eqref{eq:couplings} will be adopted. We perform a model-independent analysis of the   $pp\to jj H^\pm$, $H^\pm\to tb$ with leptonic decay of top quarks at the 13~TeV LHC. The charged Higgs boson mass is assumed to be in the range from 200~GeV to 1~TeV~\footnote{We choose this mass range to match the current searches in the VBF channels in Refs.~\cite{Sirunyan:2017sbn,Aad:2015nfa}.}. We note that in Ref.~\cite{Asakawa:2006gm}, the process $pp\to jjH^\pm$, $H^\pm\to tb$ was also studied at the LHC.  Only   the effective coupling $H^\pm W^\mp Z$ was discussed, whereas how $H^\pm$ can simultaneously decay into $tb$ was not explained~\cite{Asakawa:2006gm}. We will perform a more detailed collider analysis in this section and investigate a realistic model, which has both tree-level couplings of $H^\pm W^\mp Z$ and $H^\pm tb$ in Section~\ref{sec:model}.

We generate the signal and backgrounds $t\bar{t}$, $tW$ and $tq$ processes with $q=j,b$, using~\texttt{MG5\_aMC@NLO v2.4.3}~\cite{Alwall:2014hca} at the parton level with following cuts to isolate objects:
\begin{align}
\label{eq:basic_cuts}
p_{T}^{j,b}>20\gev,p_{T}^{\ell}>10\gev,\ \Delta R_{mn}>0.2,\ |\eta_{j,b}|<5,\ |\eta_{e,\mu}|<2.5. 
\end{align}
In the above,  $m,n=j,b,\ell$, $j$ denotes light-flavor quarks, $\ell=e,\mu$, and the angular distance in the $\eta-\phi$ plane is defined as $\Delta R_{ij}\equiv\sqrt{(\eta_i-\eta_j)^2+(\phi_i-\phi_j)^2}$ with $\eta_i$ and $\phi_i$ being the pseudo-rapidity and azimuthal
angle of particle $i$, respectively. The NN23LO1 Parton Distribution Function (PDF) set~\cite{Ball:2012cx} and default hadronization and factorization scales are used. The parton-level events are interfaced to \texttt{Pythia 6.4}~\cite{Sjostrand:2006za} and \texttt{Delphes3 v3.3.3}~\cite{deFavereau:2013fsa} for parton shower and detector simulation. The backgrounds $t\bar{t}$, $tW$, $tq$ are matched  in 5-flavor scheme up to 1, 1 and 2 jets, respectively. Jets are clustered via the anti-$k_t$ algorithm~\cite{Cacciari:2008gp} with a radius parameter of $R=0.4$ as implemented in \texttt{Fastjet}~\cite{Cacciari:2011ma}.

To select the signal process, we impose   a series of selection cuts, which are composed of basic, VBF and optimal cuts. The \textit{basic cuts} are used to identify objects at the hadron level:
\begin{enumerate}[(B-1)]
\item The angular separation is chosen as $\Delta R_{mn}>0.4$, $m,n=j,b,\ell$.
\item At least four jets with $p_T>25\gev$ are tagged, and two or more of them are $b$-tagged. It is assumed that the $b$-tagging efficiency is 0.7, while the $c$ quark and light flavor quark misidentification probability are 0.2 and 0.01, respectively~\cite{Aad:2015typ,CMS:2016kkf}.
\item Events with exactly one charged lepton with $p_T>30\gev$ and no other charged lepton with $p_T>10\gev$ are selected.	
\item It is required that the missing energy $E_T^{\text{miss}}>30\gev$~\cite{Aaboud:2018jux} since there is a neutrino of the signal process in the final state.
\end{enumerate}

The \textit{VBF cuts} are adopted similar with those in Ref.~\cite{Sirunyan:2017sbn,Aad:2015nfa}:
\begin{enumerate}[(V-1)]
\item There are at least two non-$b$-tagged jets in opposite detector hemispheres.
\item The invariant mass and rapidity separation of the leading $p_T$ jets in opposite hemispheres are used as $m_{jj}>400\gev$, $|\Delta\eta_{jj}|>3.5$.
\end{enumerate}

\begin{figure}[!htb]
\centering
\includegraphics[width=0.4\textwidth]{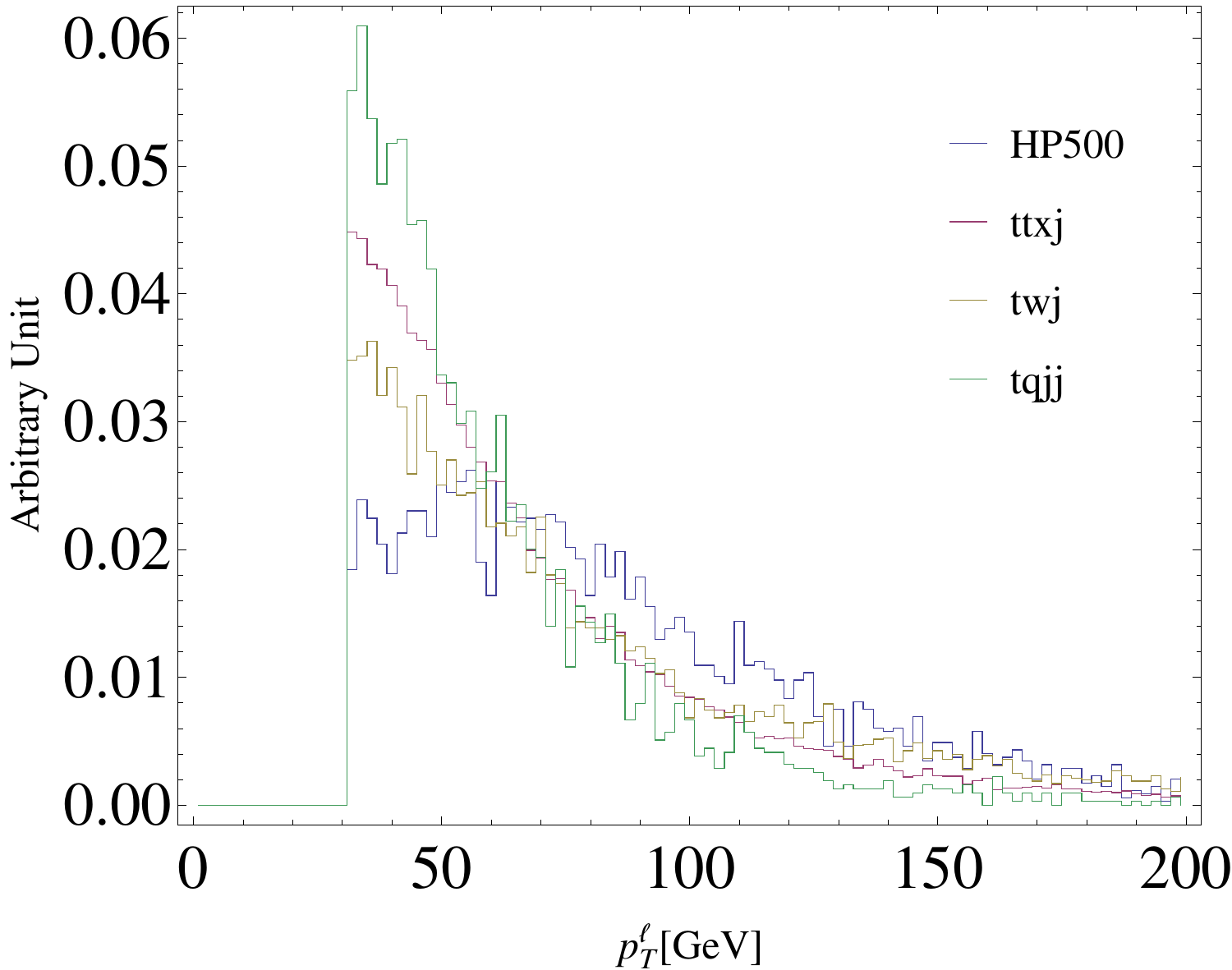}	
\includegraphics[width=0.42\textwidth]{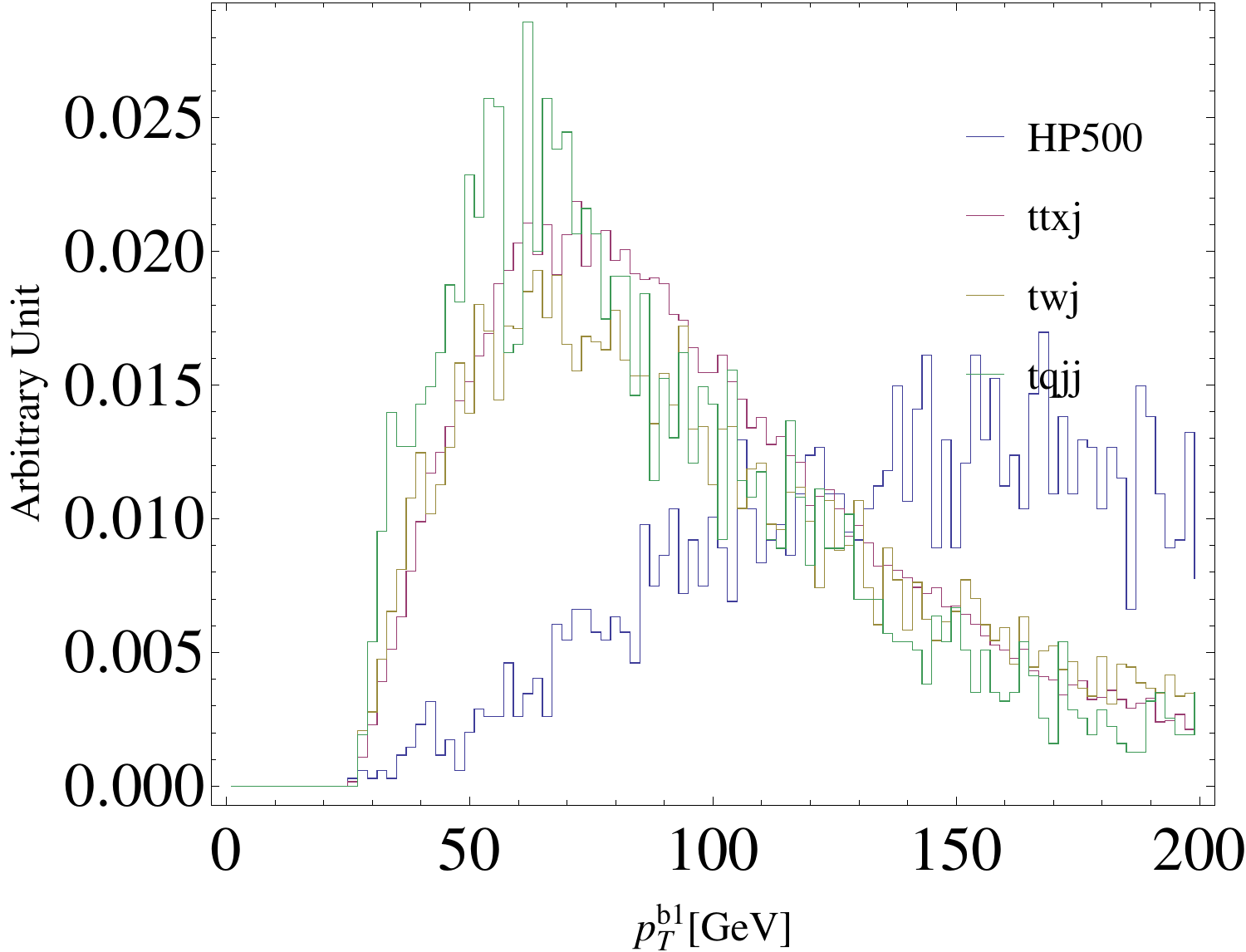}	\\
\includegraphics[width=0.4\textwidth]{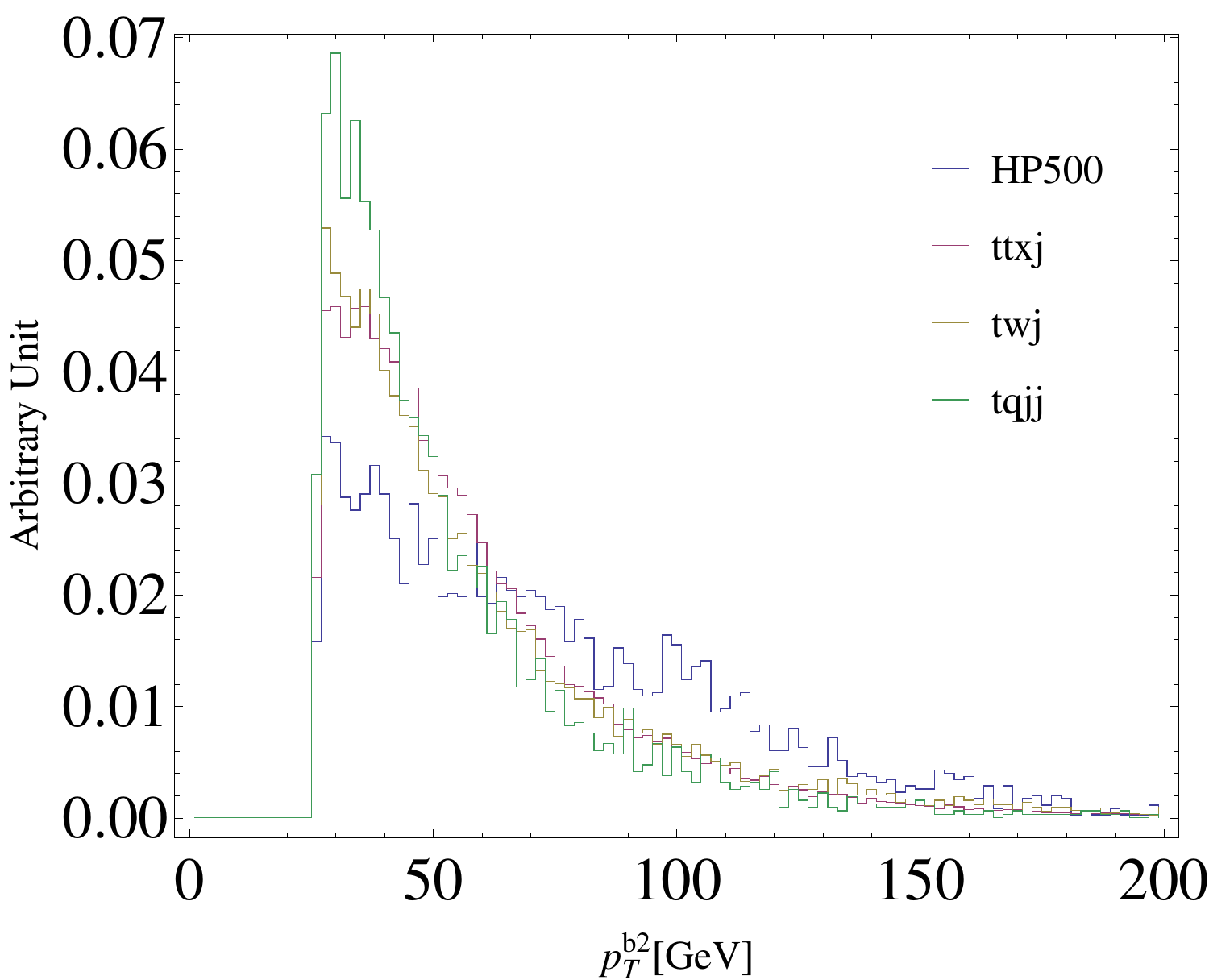}	
\includegraphics[width=0.4\textwidth]{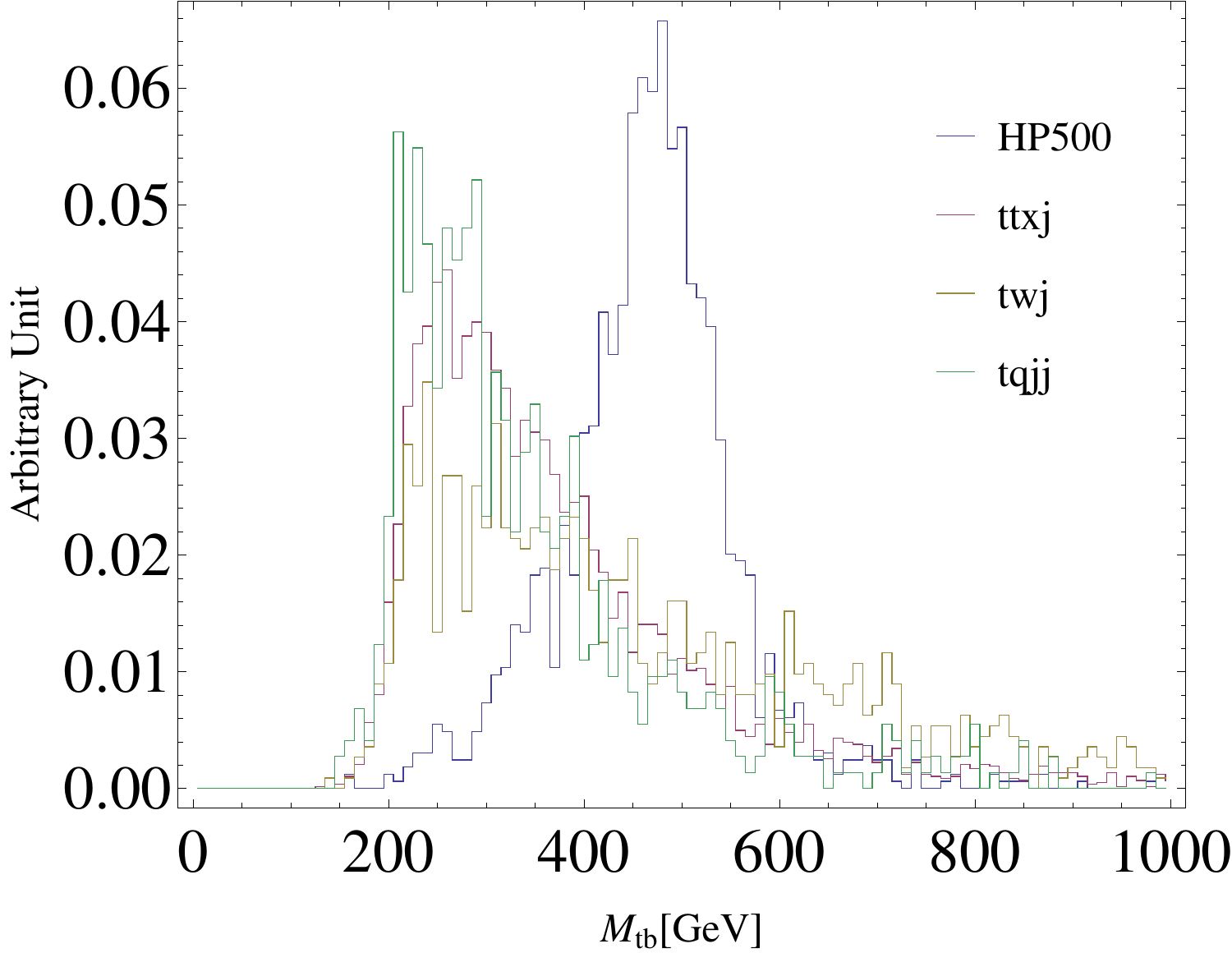}	
\caption{Kinematic distributions of $p_{T}^{\ell}$, $p_{T}^{b1}$, $p_{T}^{b2}$ and $m_{tb}$ after the VBF cuts for the charged Higgs boson mass $m_{H^\pm}=500\gev$. Here ``HP500'' and ``ttxj'', ``twj'', ``tqjj'' denote the signal and backgrounds $t\bar{t}$, $tW$, $tq$, respectively. }
\label{fig:kinematics}
\end{figure}

\begin{table}[!htb]
\tabcolsep=8pt
\caption{Bounds of kinematic variables  in the mass range  $200\gev$ $\leq m_{H^\pm}\leq$ $1000\gev$.  All numbers are in units of GeV. The symbol ``$-$'' indicates no cut being imposed. }
\begin{tabular}{c|c|c|c|c|c}
\hline
\hline
$m_{H^\pm}$ & $p_{T\text{min}}^{\ell}$ 
& $p_{T\text{min}/\text{max}}^{b1}$
& $p_{T\text{min}/\text{max}}^{b2}$
& $m_{tb}^{\text{min}}$& $m_{tb}^{\text{max}}$\\ \hline
200 & $-$ & 70 & 40 & $-$ & 250\\
225 & $-$ & 75 & 50 & $-$ & 250\\
250 & $-$ & $-$ & $-$ & $-$ & 300\\
275 & $-$ & $-$ & $-$ & $-$ & 350\\
300 & $-$ & $-$ & $-$ & $-$ & 350\\
325 & $-$ & 70 & $-$ & 250 & 350\\
350 & $-$ & 80 & $-$ & 250 & 400\\
375 & $-$ & 90 & $-$ & 275 & 450\\
400 & $-$ & 100 & $-$ & 300 & 500\\
450 & $-$ & 110 & 60 & 350 & 550\\
500 & 50 & 120 & 65 & 400 & 600\\
550 & 70 & 130 & 65 & 450 & 650\\
600 & 75 & 140 & 70 & 450 & 700\\
700 & 75 & 150 & 75 & 500 & 800\\
800 & 85 & 160 & 75 & 500 & 1000\\
900 & 85 & 170 & 80 & 550 & $-$\\
1000 & 90 & 170 & 80 & 600 & $-$\\
\hline
\hline
\end{tabular}
\label{tbl:xsection}
\end{table}

High-$p_T$ $b$-jets and charged lepton from the decay $H^\pm \to tb$, $t\to b\ell\nu$ can be further used to trigger the signal. Besides, since there is only one neutrino in the signal process, it is possible to fully reconstruct the four-momentum of top quark and the invariant mass of the charged Higgs boson $H^\pm$'s decay products. Depending on the kinematic distributions of $p_{T}^{\ell}$, $p_{T}^{b1}$, $p_{T}^{b2}$ and $m_{tb}$ after the VBF cuts for each $m_{H^\pm}$ (see Fig.~\ref{fig:kinematics} for distributions with $m_{H^\pm}=500\gev$), one can impose the \textit{optimal cuts} as follows:
\begin{enumerate}[(O-1)]
\item The transverse momentum of charged lepton
is larger than a minimum, $p_{T}^{\ell}>p_{T\text{min}}^{\ell}$.
\item Lower bounds or upper bounds of the transverse momenta of tagged $b$-jets are imposed depending on $m_{H^\pm}$,
\begin{itemize}
\item $p_{T}^{b1}<p_{T\text{max}}^{b1}$, $p_{T}^{b2}<p_{T\text{max}}^{b2}$ for $m_{H^\pm}\leq 225\gev$,
\item $p_{T}^{b1}>p_{T\text{min}}^{b1}$, $p_{T}^{b2}>p_{T\text{min}}^{b2}$ for $m_{H^\pm}\geq 325\gev$,
\end{itemize}
 where $p_{T}^{b1}$ and $p_{T}^{b2}$ denote the leading and sub-leading transverse momenta of $b$-jets.  
\item The four-momentum of top quark can be reconstructed using the template $\chi^2$ method. For each event, the $\chi^2$ is defined as~\cite{Kobakhidze:2014gqa}
\begin{align}
\chi^2=\left(\dfrac{m_{b\ell\nu}-m_t}{\Gamma_t}\right)^2,
\end{align}
where $m_t=172.5\gev$~\cite{Tanabashi:2018oca}, $\Gamma_t=4.08\gev$, $m_{b\ell\nu}$ denotes the invariant mass of lepton, neutrino and one $b$-jet. With the on-shell condition of $W$ boson, we can determine the longitudinal transverse momentum of neutrino with a two-fold ambiguity~\cite{Berger:2011xk}
\begin{align}
p_{L}^{\nu}=\dfrac{1}{2(p_{T}^{\ell})^2}\bigg[ A_W p_{L}^{\ell}\pm E_{\ell}\sqrt{A_W^2-4(p_{T}^{\ell})^2(E_{T}^{\text{miss}})^2}\bigg],
\end{align}
where $A_W=m_W^2+2\vec{p}_{T}^{~\ell}\cdot \vec{p}_{T}^{~\text{miss}}$ and $m_W=80.4\gev$. Furthermore, there are two $b$-jets in the final state~\footnote{Actually, the  additional $b$-jet may  come  from the parton-level scattering with initial $b$ quark. However, such contribution should be  small  since the $b$-quark  PDF is small. To reconstruct the  top quark, we select the $b$-jet  among the leading-$p_T$ and sub-leading $p_T$ $b$-jets.}. By minimizing $\chi^2$~\cite{Aad:2014xea}, we can determine which $b$-jet originates from top quark decay and the sign of $p_{L}^{\nu}$. Thus one is able to fully reconstruct the four-momentum of top quark and the invariant mass of $t$ and $b$ $(m_{tb})$. Since in the signal process, $t$ and $b$ come from the decay of on-shell charged Higgs boson $H^\pm$, the variable $m_{tb}$ can be used  to suppress the backgrounds. The cut on $m_{tb}$ is optimized depending on $m_{H^\pm}$,
\begin{itemize}
\item  $m_{tb}^{\text{min}}<m_{tb}<m_{tb}^{\text{max}}$.
\end{itemize}
\end{enumerate}
Here, $p_{T\text{min}}^{\ell}$ denotes the lower bound of $p_{T}^{\ell}$. Similarly, $p_{T\text{min/max}}^{b1}$, $p_{T\text{min/max}}^{b2}$ and $m_{tb}^{\text{min/max}}$ are the lower/upper bounds of $p_{T}^{b1}$, $p_{T}^{b2}$ and $m_{tb}$, respectively. The explicit values are given in Tab.~\ref{tbl:xsection}.

\begin{figure}[!htb]
\centering
\includegraphics[width=0.3\textwidth]{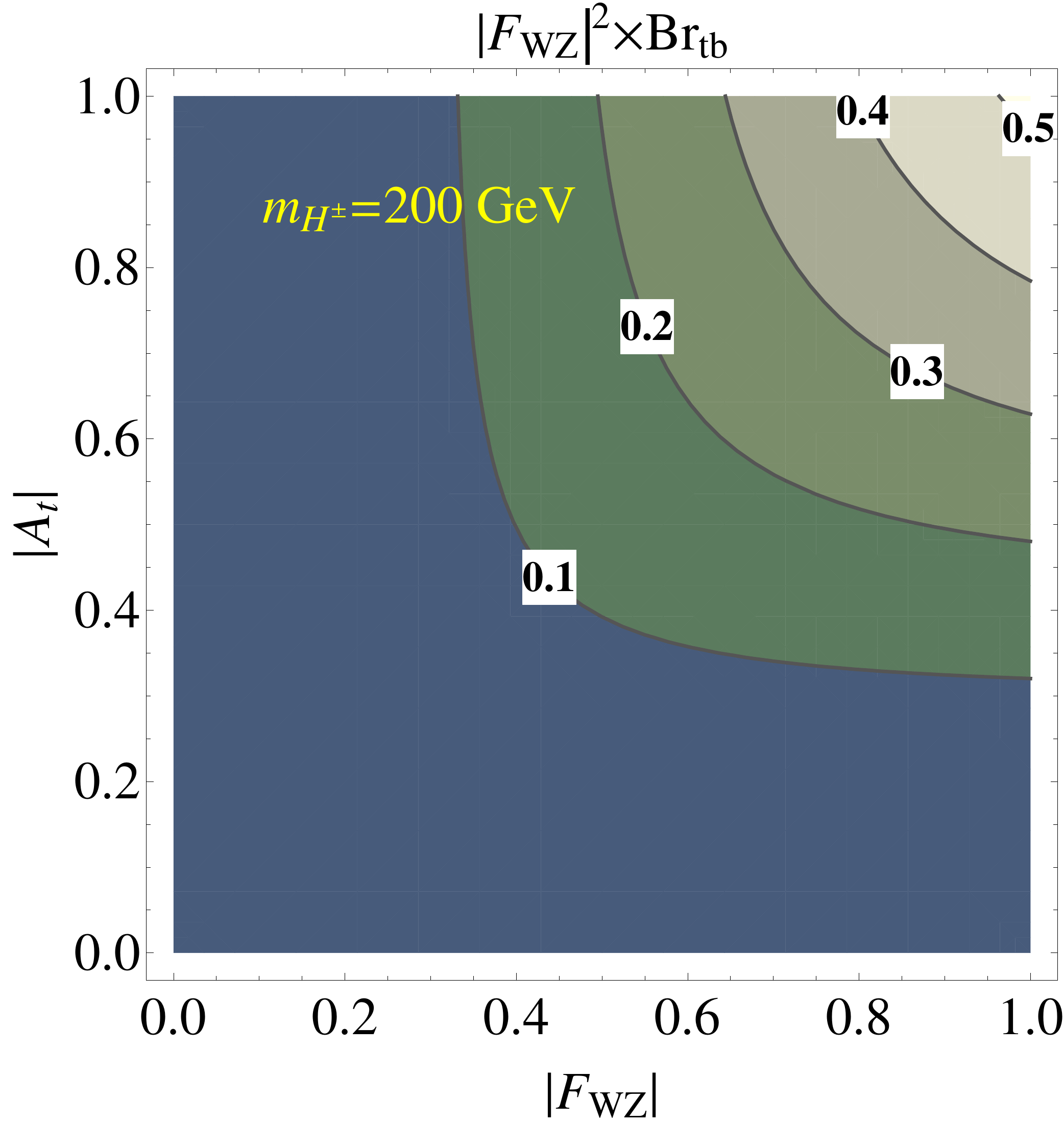}	
\includegraphics[width=0.3\textwidth]{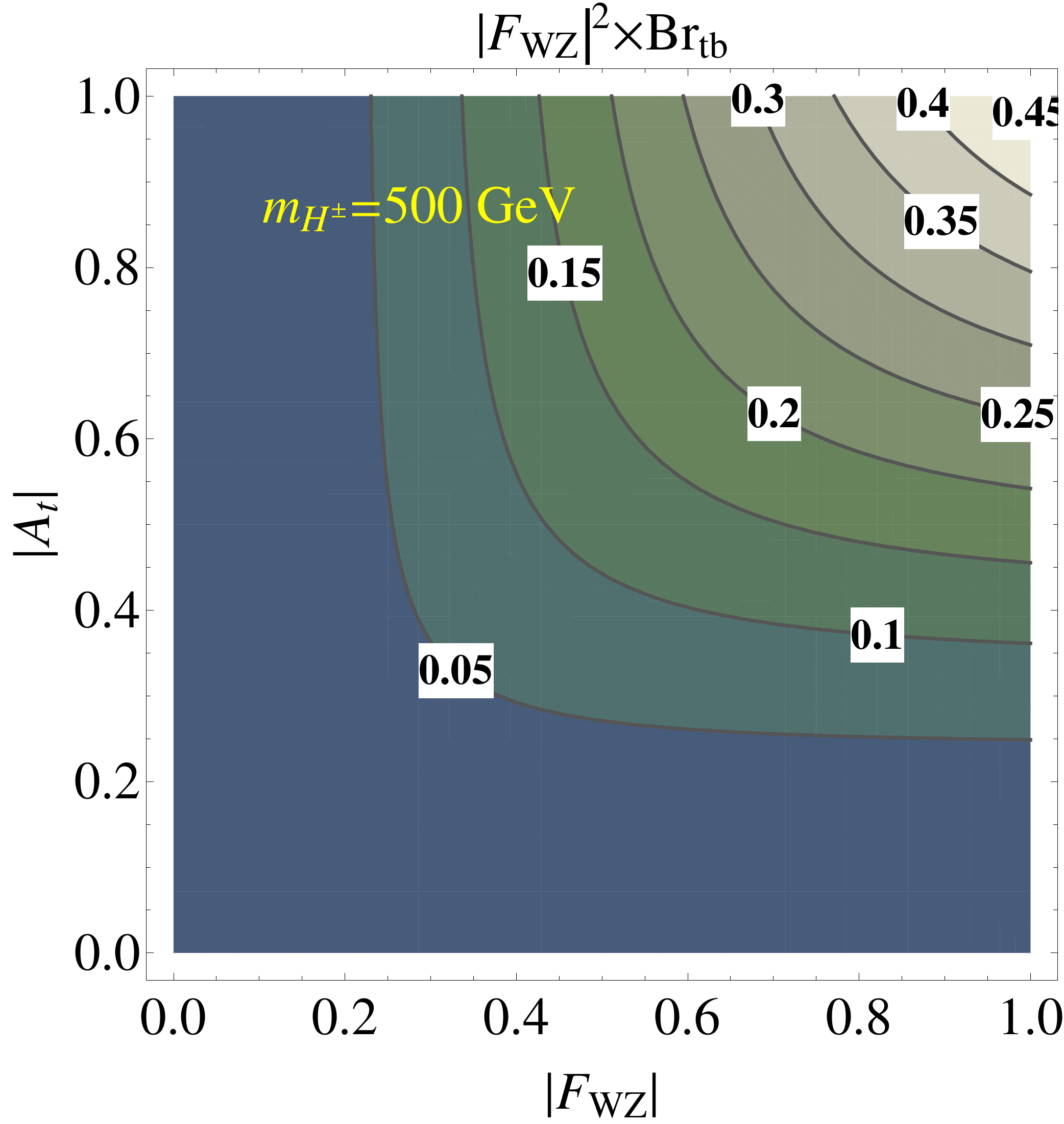}	
\includegraphics[width=0.3\textwidth]{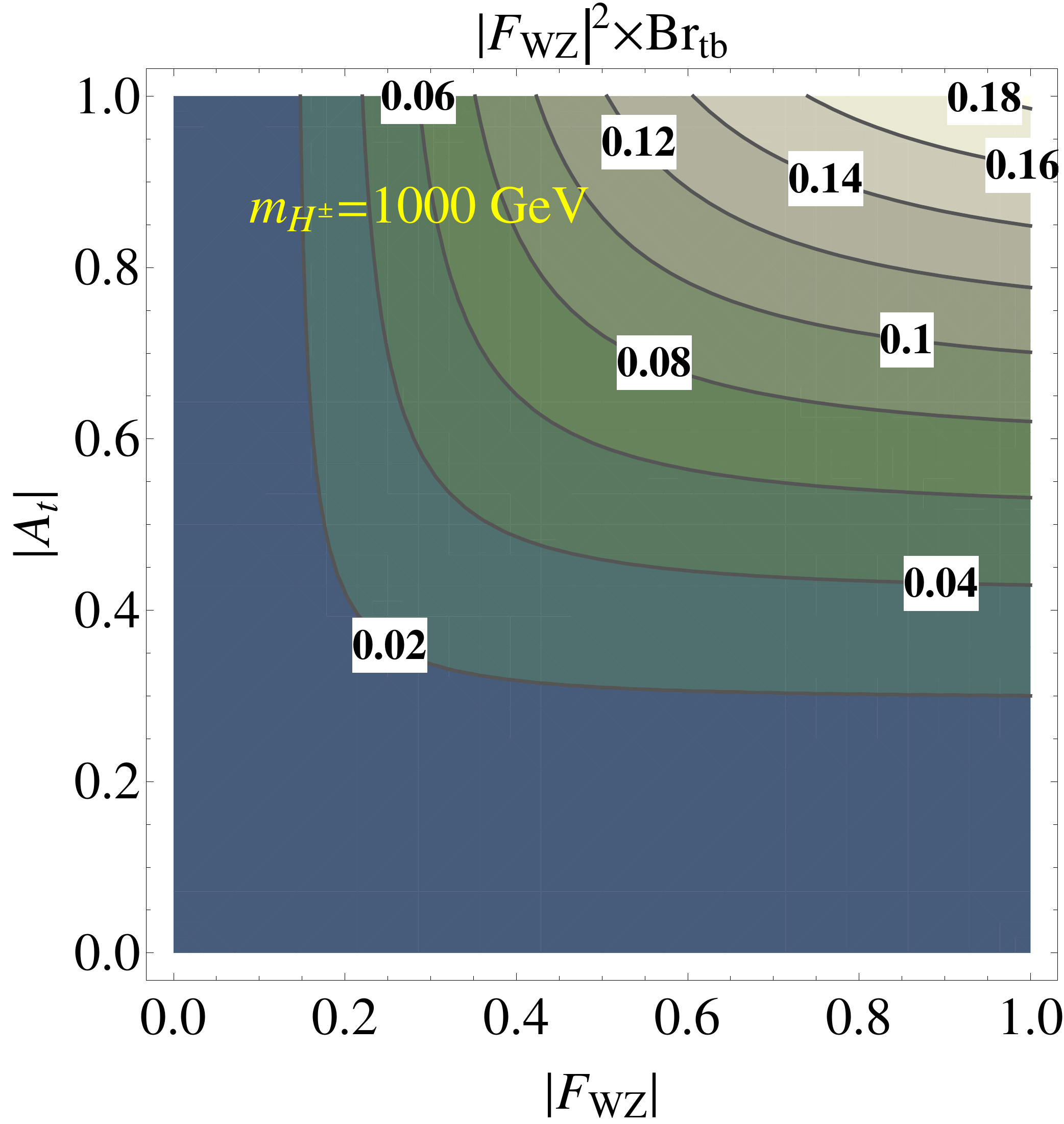}	
\caption{Dependence of $|F_{WZ}|^2\times \mathcal{B}_{tb}$ on $|F_{WZ}|$ and $|A_{t}|$ for $m_{H^\pm}=200,$ $500$, $1000\gev$. }
\label{fig:dependence} 
\end{figure}

After imposing  the selection cuts, we find that the significance of the signal can be greatly improved. We show in Tab.~\ref{tbl:cut-flow} the cut flow of the signal and background cross sections after each selection cut for $m_{H^\pm}=500\gev$ with the benchmark scenario $F_{WZ}=A_t=0.5$ . The discovery prospect and exclusion limit of the signal process are evaluated using~\cite{Cowan:2010js}
\begin{align}
\label{eq:discovery}
\mathcal{Z}_D&=\sqrt{2\bigg [(n_s+n_b)\log\dfrac{n_s+n_b}{n_b}-n_s\bigg ]},\\
\label{eq:exclusion}
\mathcal{Z}_E&=\sqrt{-2\bigg [n_b\log\dfrac{n_s+n_b}{n_b}-n_s\bigg ]},
\end{align}
respectively.  $n_s=\sigma_s\mathcal{L}$ and $n_b=\sigma_b\mathcal{L}$ denote the number of events after cuts with the integrated luminosity of $\mathcal{L}$, $\sigma_s$ and $\sigma_b$ are the cross sections of the signal and total background after cuts. The signal cross section $\sigma_s$ can be expressed as
\begin{align}
\label{eq:mixed}
\sigma_s&=\sigma(pp \to jjH^\pm)\mathcal{B}(H^\pm\to tb)\epsilon_{s},\nn\\
&=\big[\sigma(pp \to jjH^\pm)_{\text{BM}}\mathcal{B}(H^\pm\to tb)_{\text{BM}}\epsilon_{s}\big]\dfrac{|F_{WZ}|^2}{0.5^2}\dfrac{\mathcal{B}(H^\pm\to tb)}{\mathcal{B}(H^\pm\to tb)_{\text{BM}}},
\end{align}
where the quantities in the square bracket are obtained  with the benchmark values $F_{WZ}=A_t=0.5$. The signal cut efficiency $\epsilon_s$ is independent of the couplings $F_{WZ}$ and $A_t$~\footnote{We assume that the narrow width approximation is always valid for the signal process. For $|F_{WZ}|,|A_t|\lesssim 1$ and $m_{H^+}\leq 1000\gev$,  we obtain   $\Gamma_{H^+}/m_{H^+}\lesssim 30\%$.}, so that we are free to obtain it with the benchmark values of $F_{WZ}$ and $A_t$. The signal cross section is proportional to $|F_{WZ}|^2\times \mathcal{B}_{tb}$, which depends on $|F_{WZ}|$ and $|A_t|$ as shown in Fig.~\ref{fig:dependence}. We find that $|F_{WZ}|^2\times \mathcal{B}_{tb}$ tends to be larger for a smaller $m_{H^\pm}$.

\begin{table}[!htb]
\tabcolsep=8pt
\caption{The cut flow of the signal and background cross sections (in units of pb) for $m_{H^\pm}=500\gev$. The notation of ``aE$\pm$0b'' stands for $a\times 10^{\pm b}$.}
\begin{tabular}{c|cccc}
\hline
\hline
cuts  & signal & $t\bar{t}$ & $tW$ & $tq$\\ \hline
cuts in Eq.~\eqref{eq:basic_cuts}	&	7.76E-03	&	9.96E+01	&	1.04E+01	&	3.02E+01	\\
$\Delta R_{mn} > 0.4$	&	7.76E-03	&	9.96E+01	&	1.04E+01	&	3.02E+01	\\
$n_j\geq 4$	&	6.53E-03	&	8.06E+01	&	5.67E+00	&	4.16E+00	\\
$b$-tagging 	&	3.23E-03	&	3.14E+01	&	1.53E+00	&	1.28E+00	\\
single lepton 	&	2.03E-03	&	1.50E+01	&	7.97E-01	&	5.02E-01	\\
$E_T^{\text{miss}}>30\gev$	&	1.62E-03	&	1.15E+01	&	6.12E-01	&	3.70E-01	\\
$\geq 2$ non-$b$ jets	&	1.35E-03	&	6.19E+00	&	3.12E-01	&	1.77E-01	\\
$|\Delta \eta_{jj}|>3.5$	&	1.02E-03	&	1.10E+00	&	5.35E-02	&	8.31E-02	\\
$m_{jj}>400\gev$	&	9.52E-04	&	8.41E-01	&	3.94E-02	&	5.91E-02	\\
$p_T^\ell > 50\gev$	&	7.58E-04	&	5.10E-01	&	2.76E-02	&	2.92E-02	\\
$p_T^{b1}>120\gev$, $p_T^{b2}>65\gev$	&	3.04E-04	&	9.13E-02	&	7.49E-03	&	4.78E-03	\\
$400\gev<m_{tb}<600\gev$	&	1.42E-04	&	2.33E-02	&	1.11E-03	&	7.88E-04	\\
\hline
\hline
\end{tabular}
\label{tbl:cut-flow}
 \end{table}
 
\begin{figure}[!htb]
\centering
\includegraphics[width=0.4\textwidth]{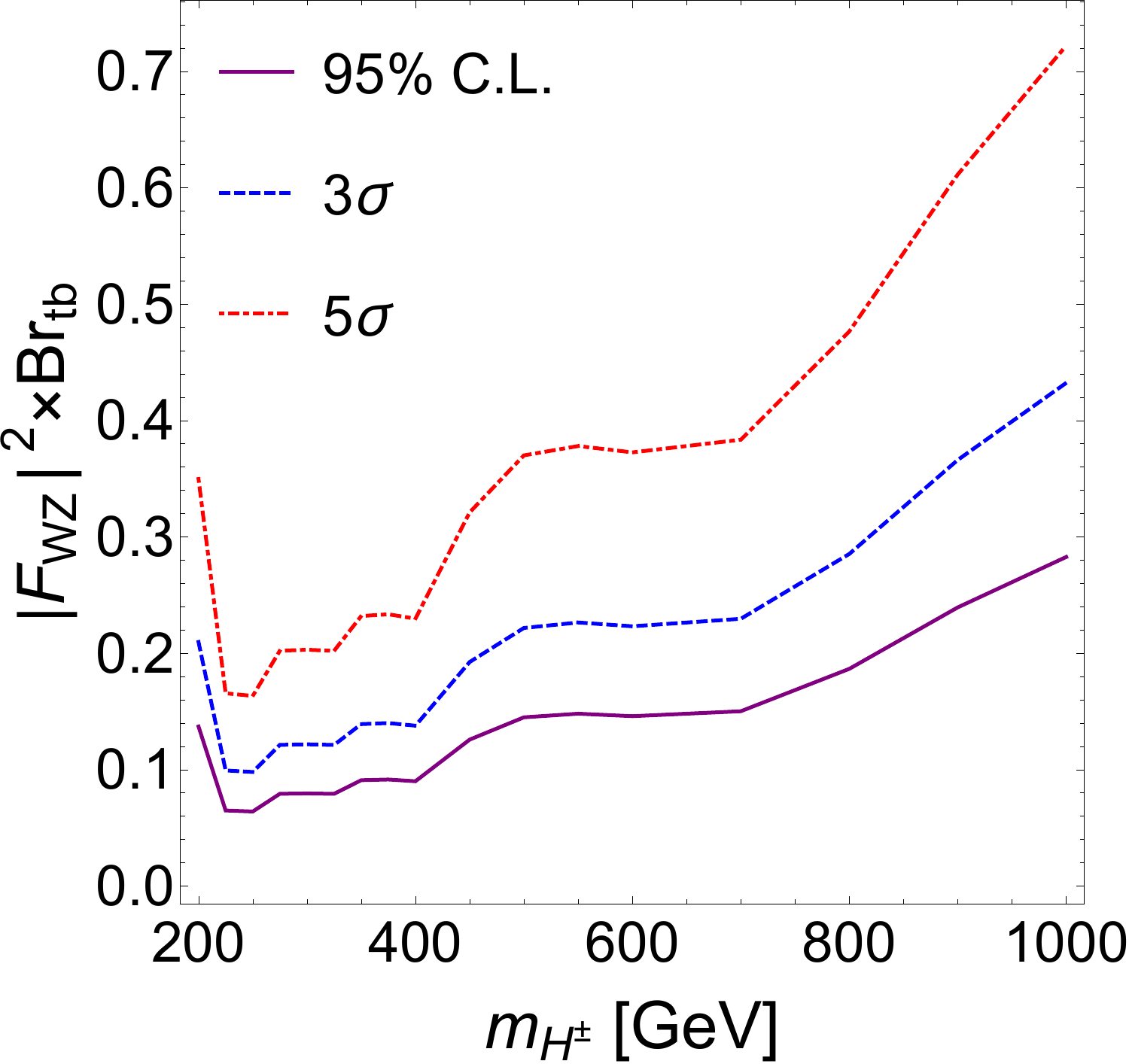}		
\caption{Sensitivities to $|F_{WZ}|^2\times \mathcal{B}_{tb}$ in $pp\to jj H^\pm,H^\pm\to tb$ with the integrated luminosity of $3\abi$. }
\label{fig:sensitivity}
\end{figure}

At the 13-14~TeV LHC, the integrated luminosity is planned to reach $3\abi$~\cite{Apollinari:2017cqg}. In this work, we will study the  process $pp\to jjH^\pm$, $H^\pm\to tb$ with integrated luminosity of $3\abi$ at the 13~TeV LHC. Using the relation in Eqs.~\eqref{eq:discovery}~\eqref{eq:exclusion}~\eqref{eq:mixed}, we can obtain sensitivity to $|F_{WZ}|^2\times \mathcal{B}(H^\pm \to tb)$ from the discovery prospects and exclusion limit of our proposed signal process $pp\to jj H^\pm$, $H^\pm \to tb$ as shown in Fig.~\ref{fig:sensitivity}.  From this figure, one can find that, the scenarios with $|F_{WZ}|^2\times \mathcal{B}(H^\pm \to tb)\gtrsim 0.43,0.72$ can be observed at $3\sigma,5\sigma$ for the charged Higgs boson in the mass range from 200~GeV to 1~TeV, respectively. While the region of $|F_{WZ}|^2\times \mathcal{B}(H^\pm \to tb)\gtrsim 0.25$ would be excluded at 95\% confidence level (C.L.) or equivalently with $\mathcal{Z}_E\geq 1.96$. It is interesting to note that the sensitivity is the best for $m_{H^\pm}\simeq 250\gev$ in the process $pp\to jjH^\pm$, $H^\pm\to tb$.

\begin{figure}[!htb]
\centering
\includegraphics[width=0.4\textwidth]{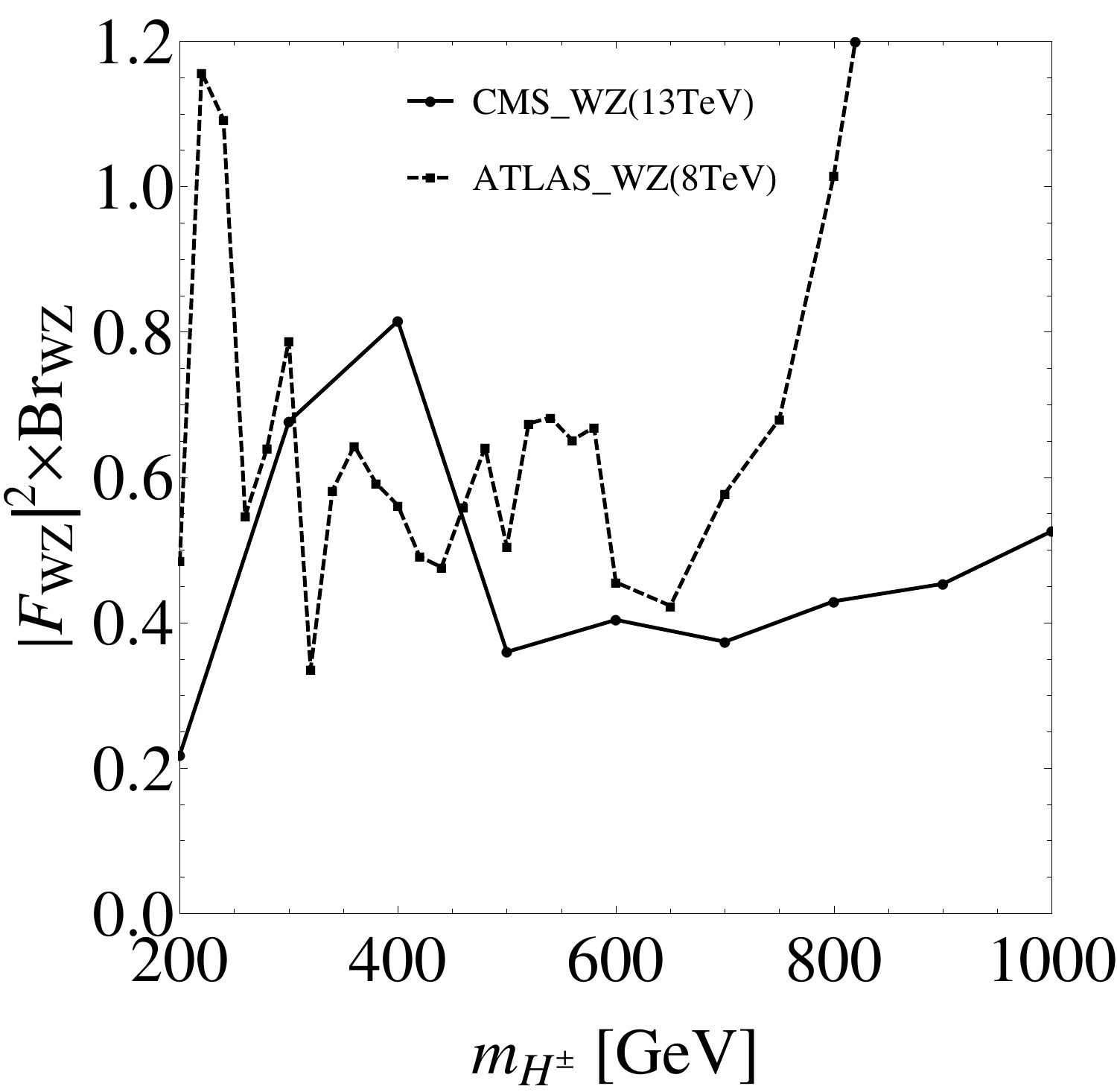}	
\includegraphics[width=0.4\textwidth]{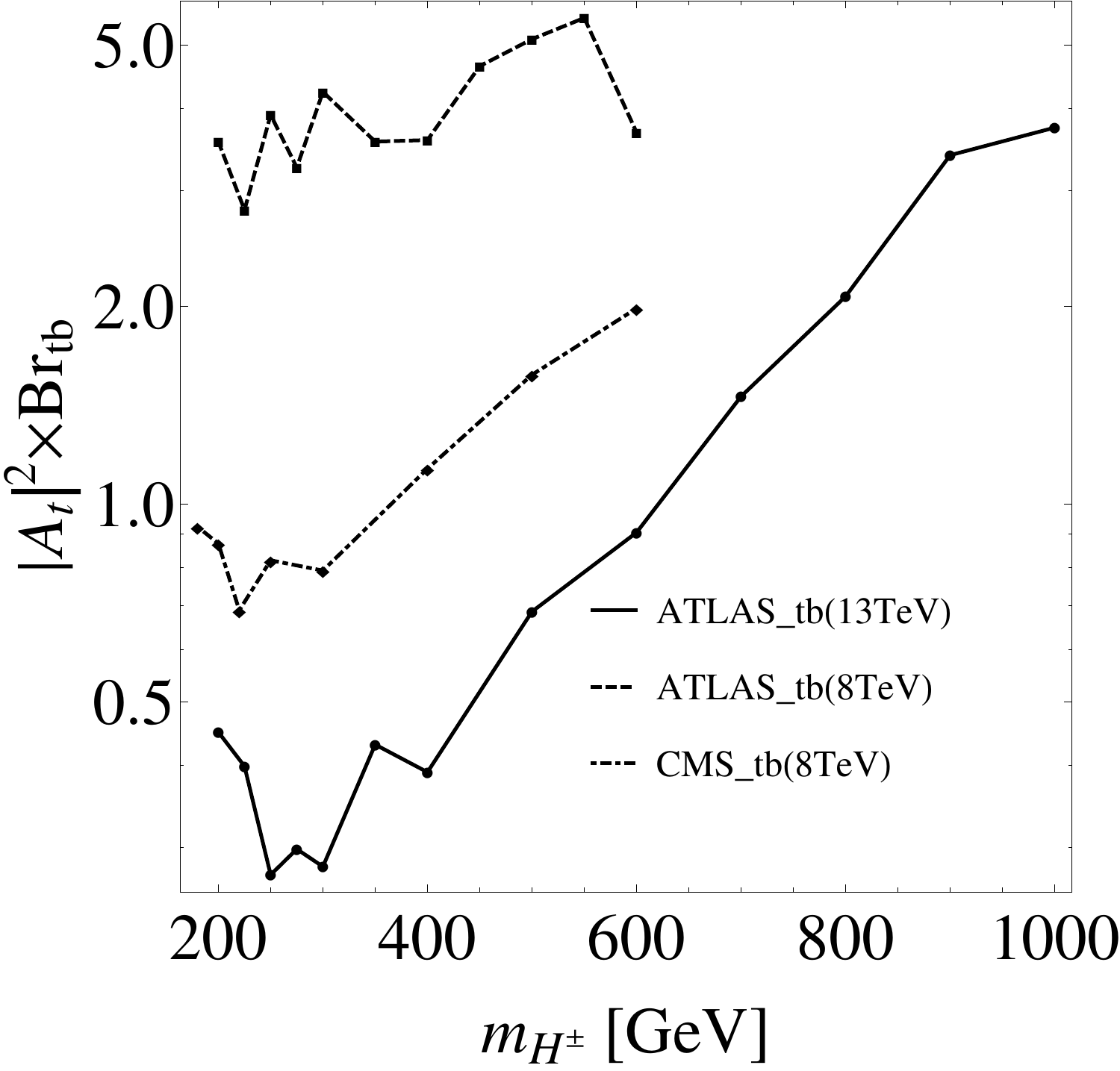}	
\caption{Constraints on $|F_{WZ}|^2\times \mathcal{B}_{WZ}$ (left panel) and $|A_{t}|^2\times \mathcal{B}_{tb}$ (right panel) from the existing charged Higgs boson searches at the LHC. }
\label{fig:constraints}
\end{figure}

\begin{figure}[!htb]
\centering
\includegraphics[width=0.3\textwidth]{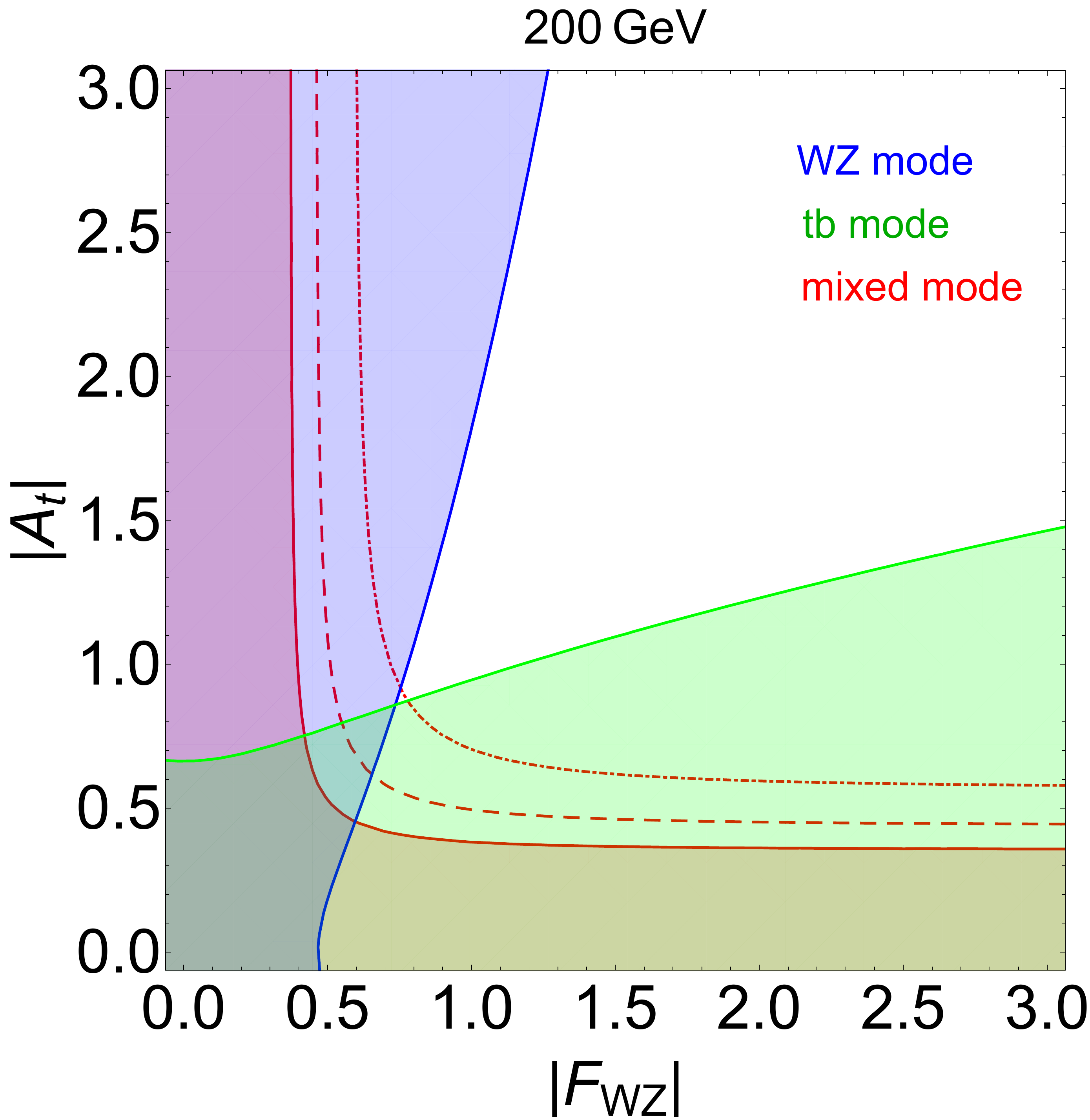}	
\includegraphics[width=0.3\textwidth]{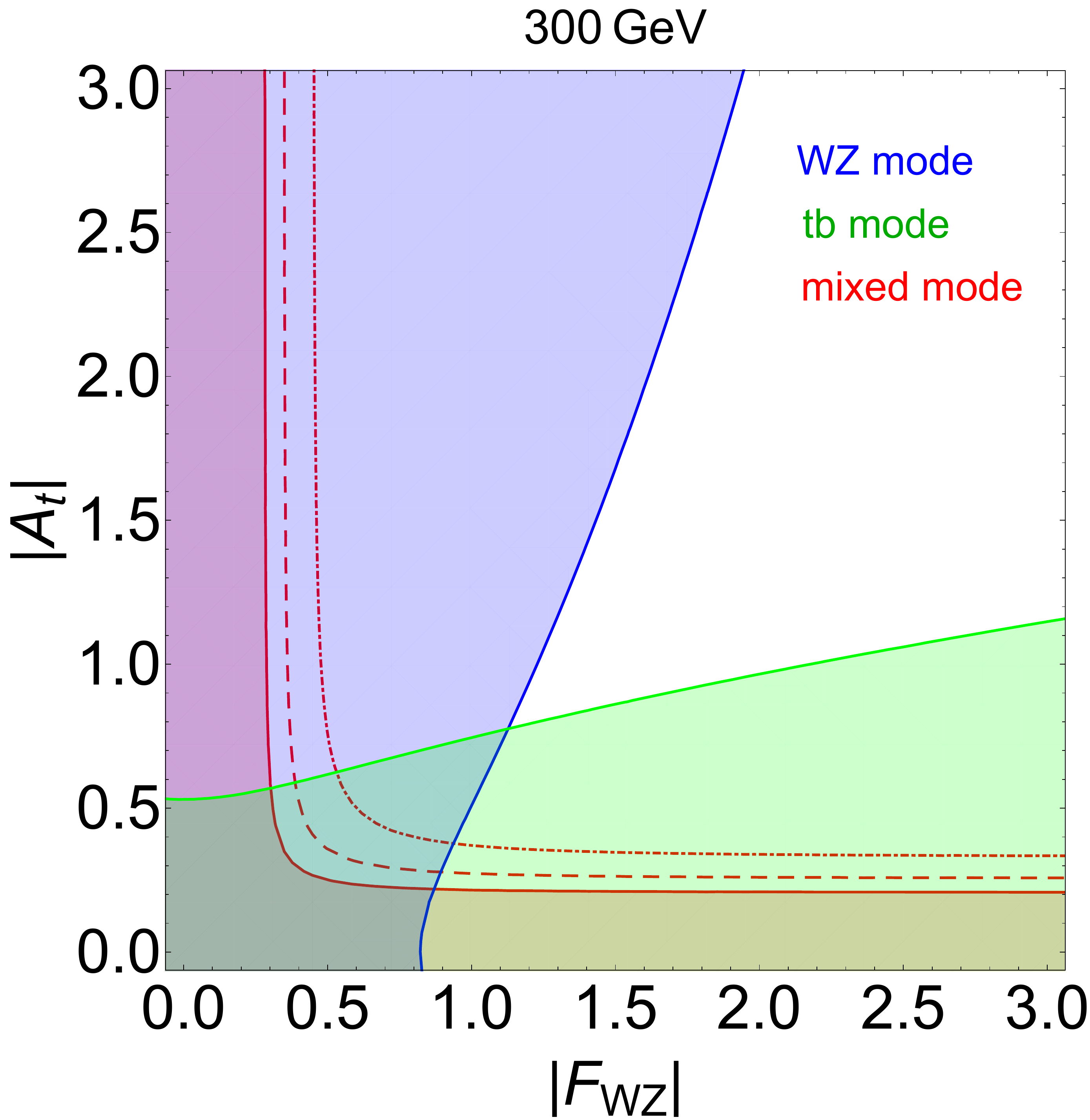}	
\includegraphics[width=0.3\textwidth]{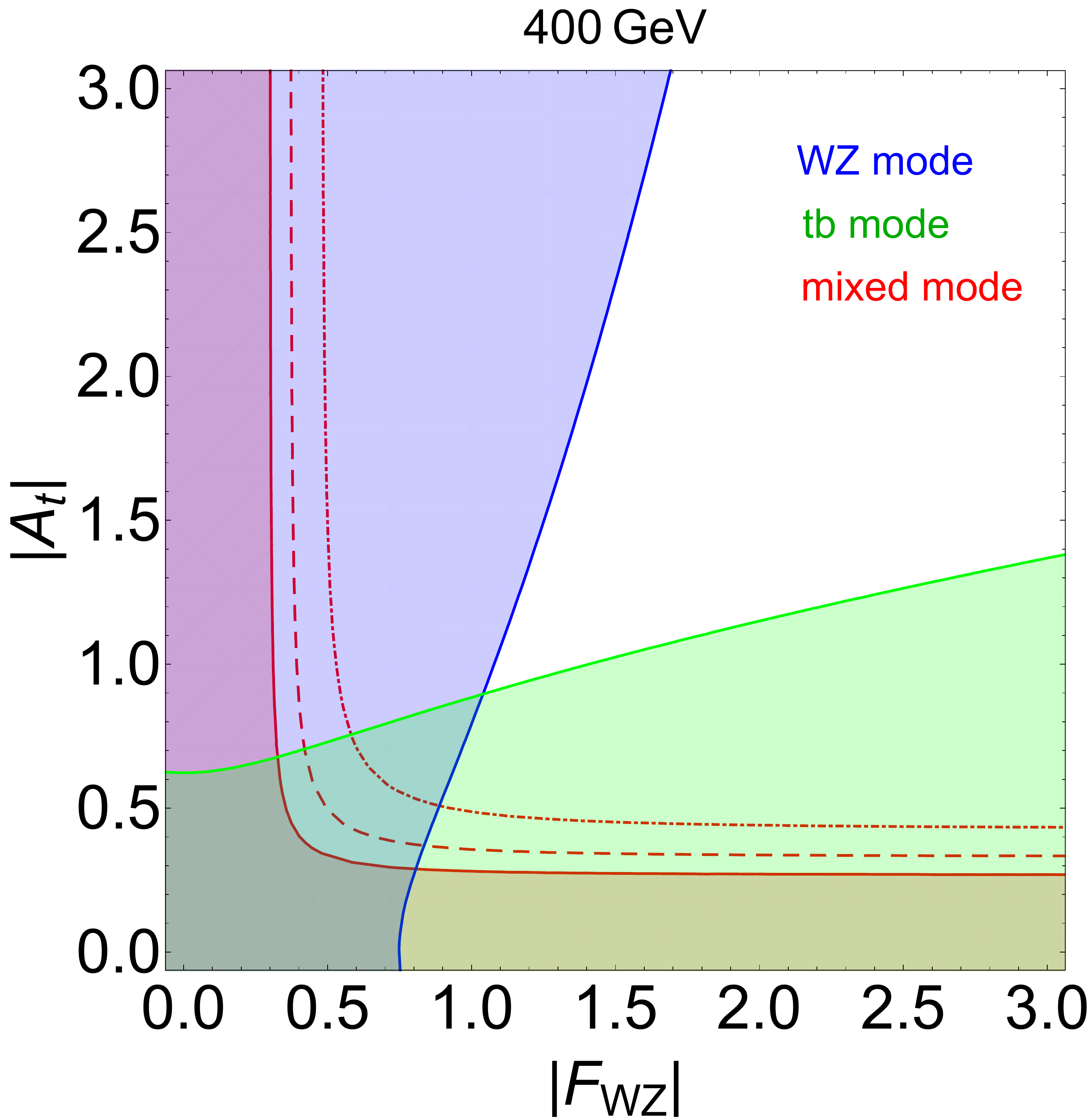}	
\includegraphics[width=0.3\textwidth]{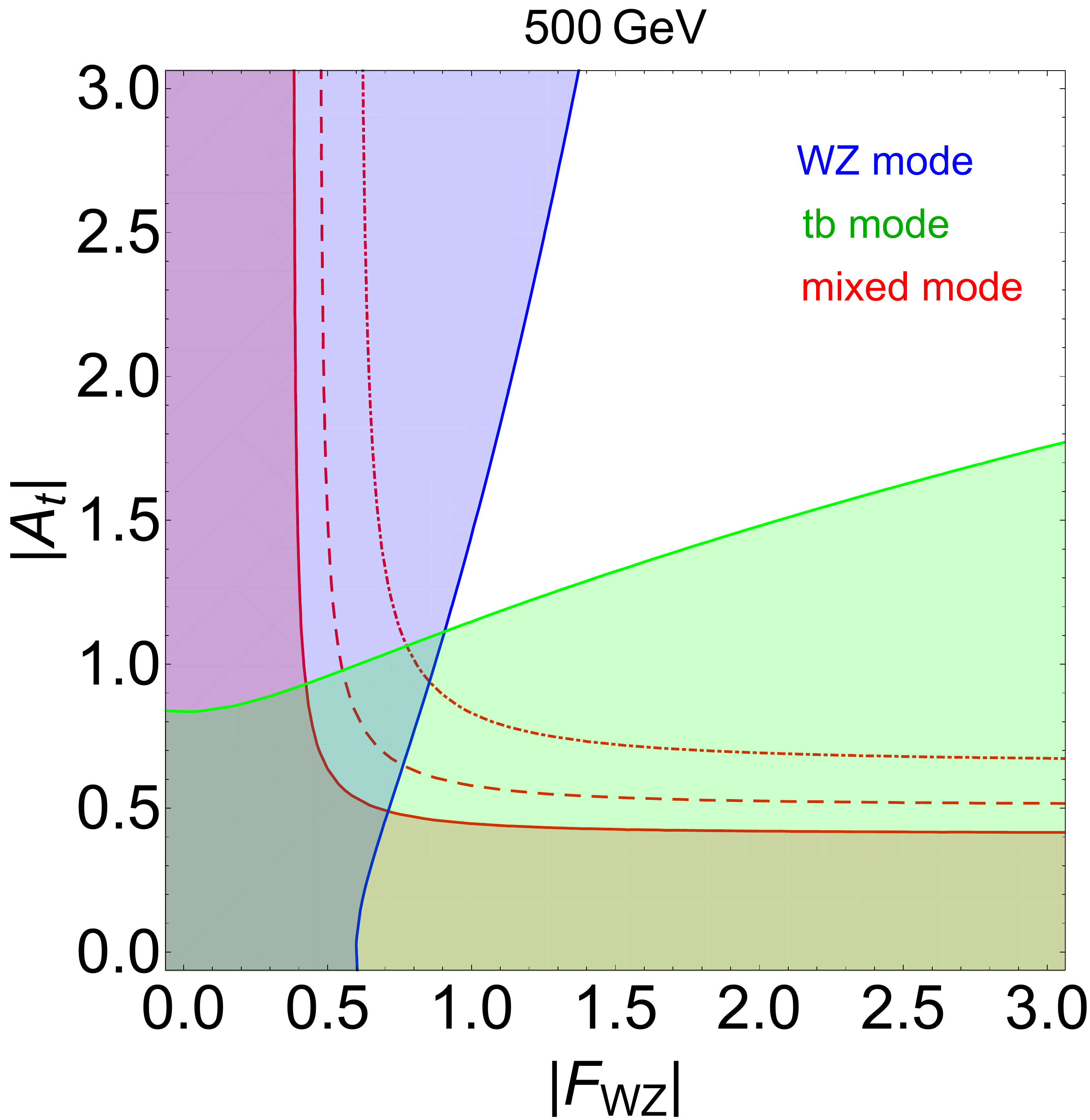}	
\includegraphics[width=0.3\textwidth]{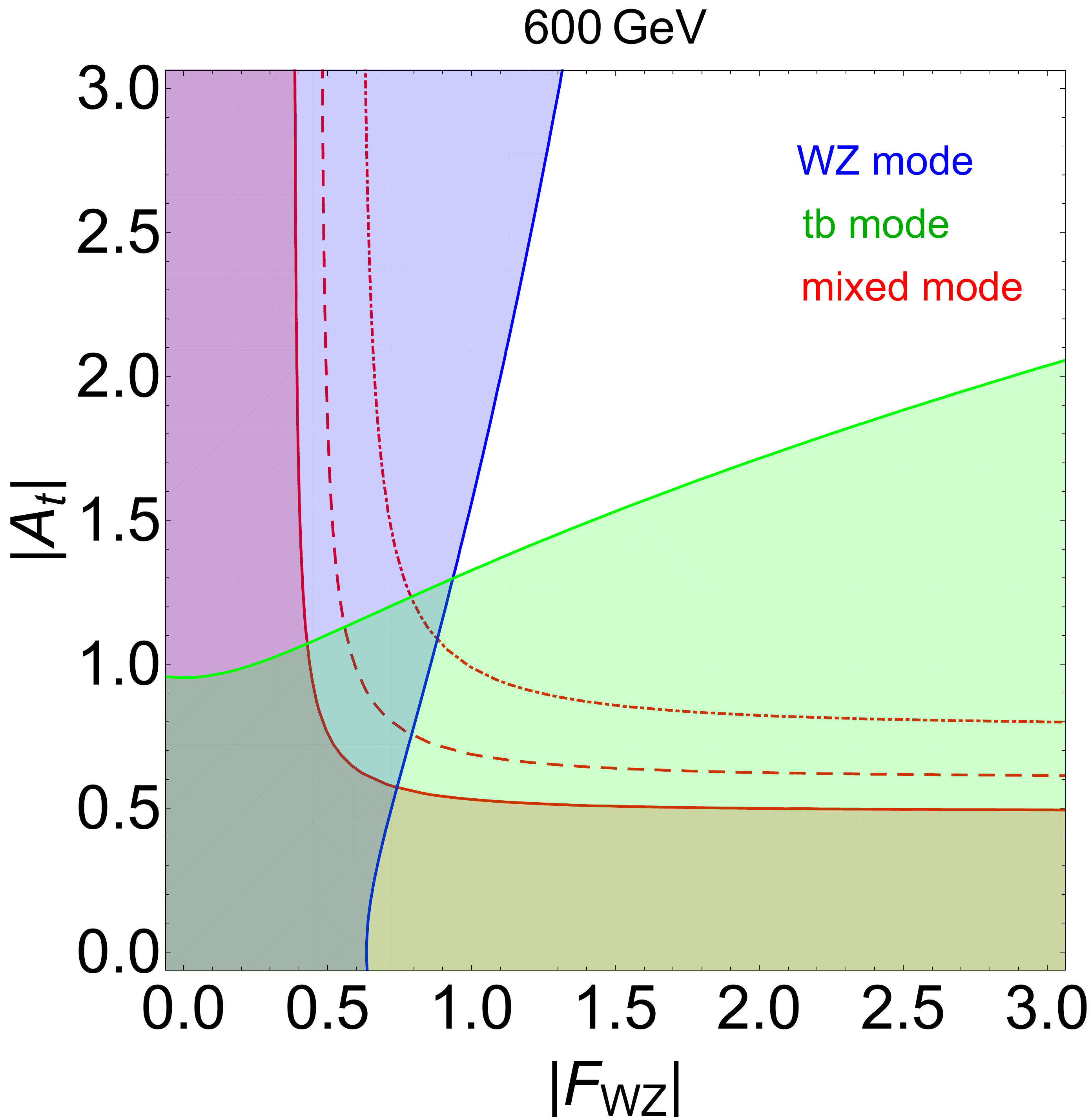}	
\includegraphics[width=0.3\textwidth]{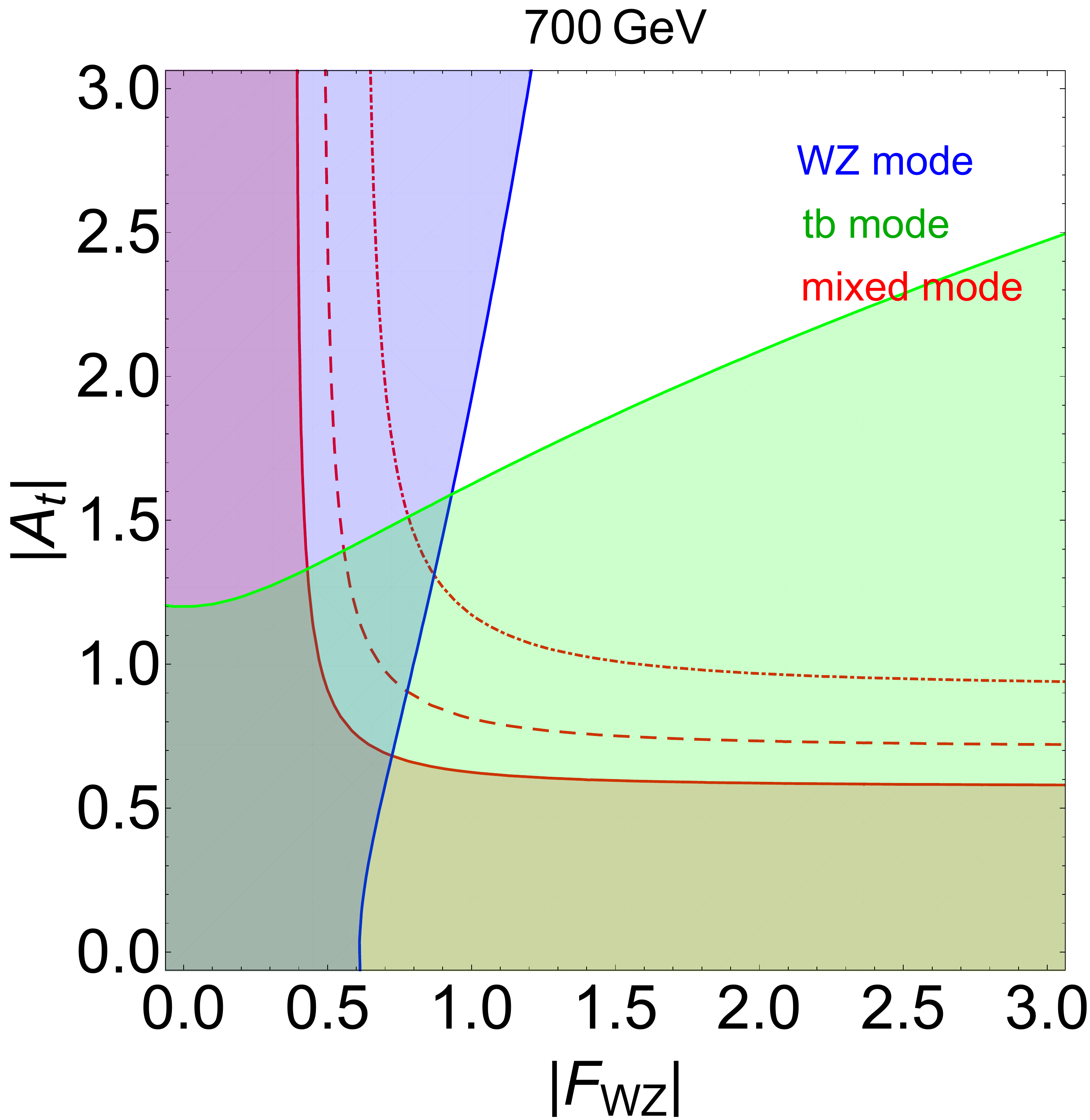}	
\includegraphics[width=0.3\textwidth]{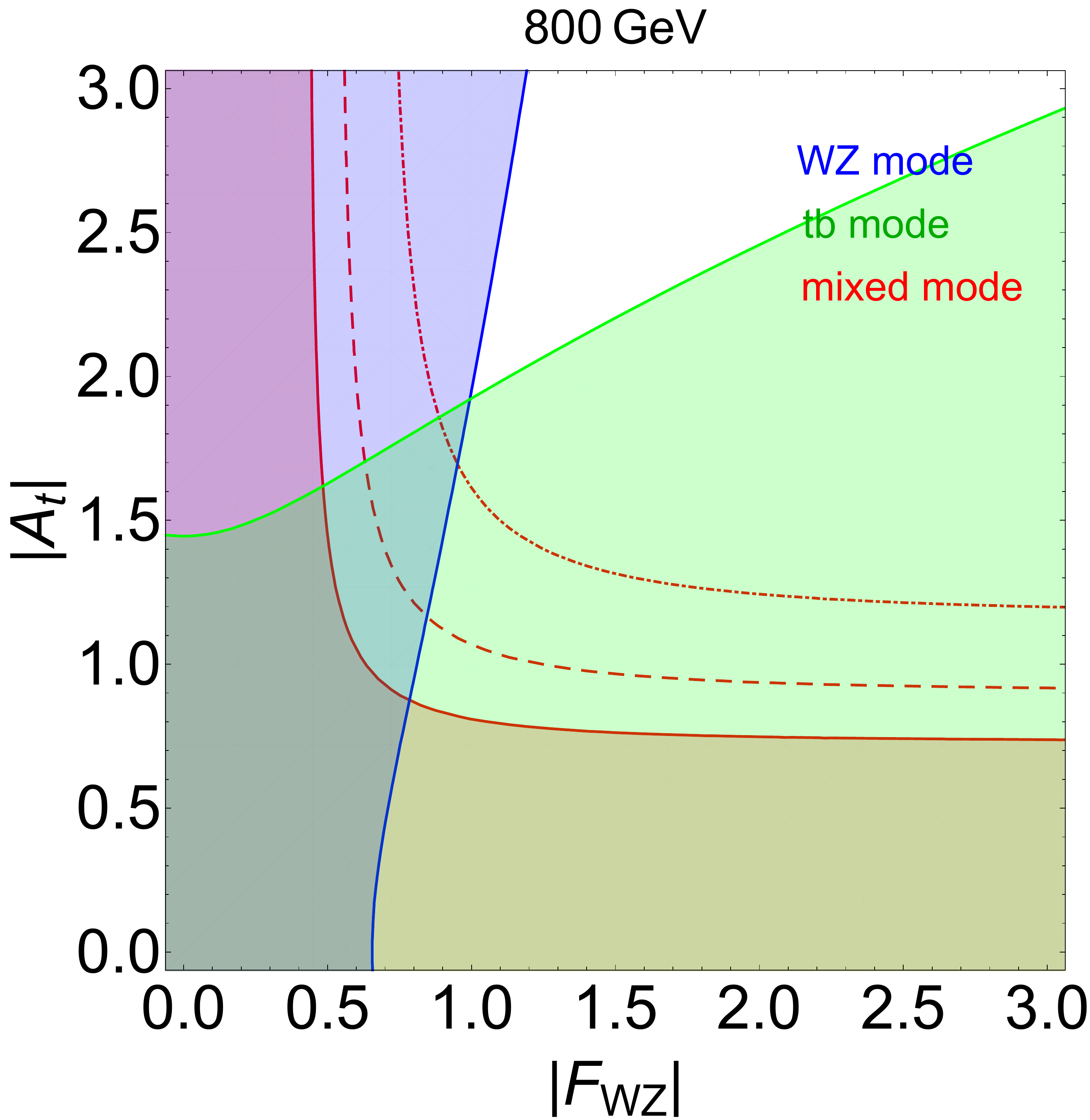}	
\includegraphics[width=0.3\textwidth]{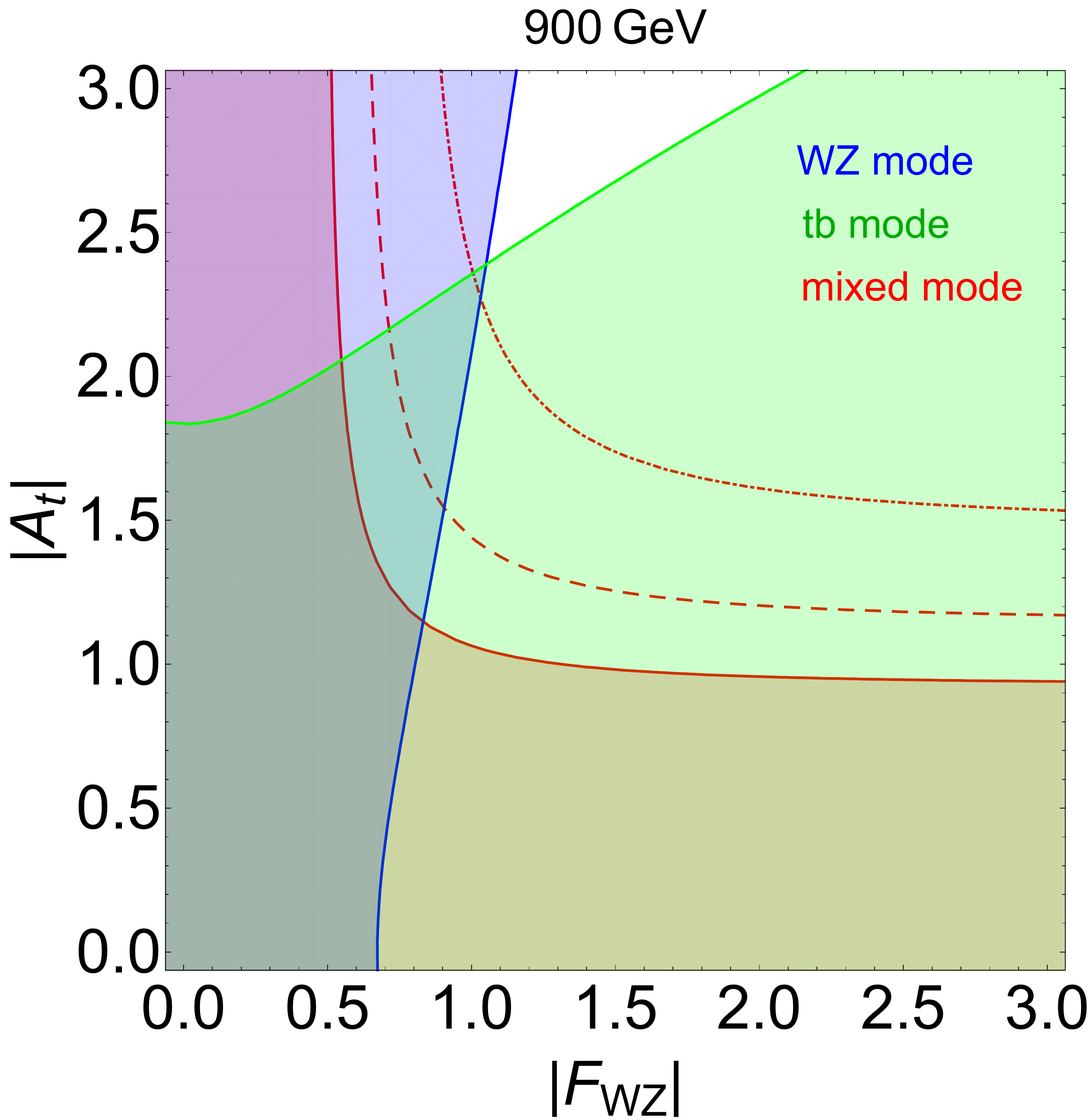}	
\includegraphics[width=0.3\textwidth]{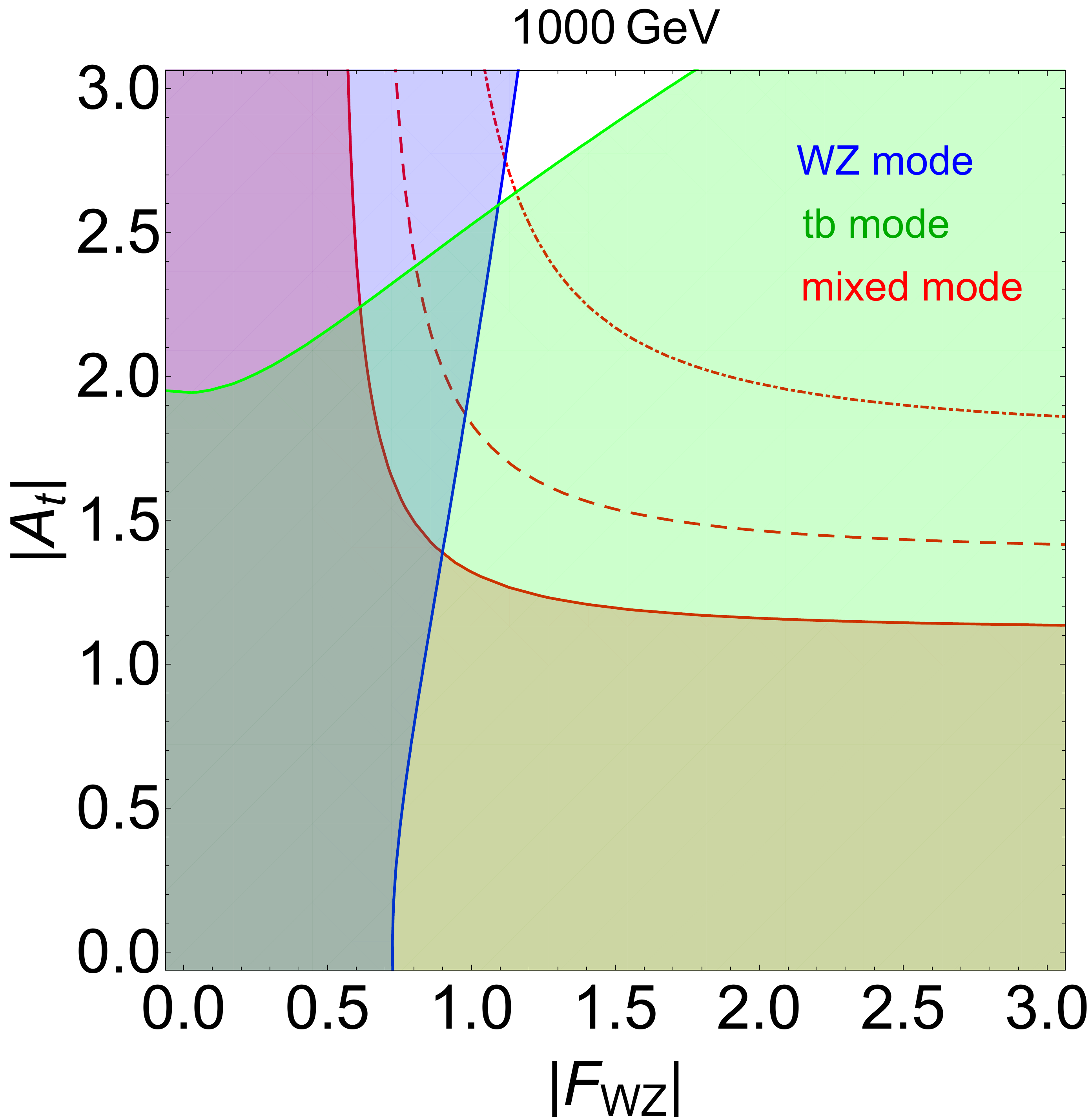}	
\caption{Sensitivities to the $H^\pm W^\mp Z$ and $H^\pm tb$ couplings for $m_{H^\pm}=200-1000\gev$ with the minimal total width assumption. The blue and green regions are obtained from the exclusion limits at 95\% C.L. of the processes $pp\to tH^\pm$, $H^\pm\to tb$ and $pp\to jjH^\pm$, $H^\pm \to W^\pm Z$, respectively. The red solid, dashed and dot-dashed curves correspond to the exclusion limits at 95\% C.L. and discovery significance $\mathcal{Z}_D=3,5$, respectively.}
\label{fig:couplings}
\end{figure}

On the other hand, there are already experimental searches for charged Higgs bosons in the 2HDMs and the GM model. They  provided constraints on  $|A_{t}|^2\times \mathcal{B}(H^\pm \to tb)$ and $|F_{WZ}|^2\times \mathcal{B}(H^\pm \to W^\pm Z)$, respectively. To recast current constraints, we generate the leading-order (LO) processes $pp\to \bar{t}H^\pm$ and $pp\to jjH^\pm$ in 5-flavor scheme using \texttt{MG5\_aMC@NLO v2.4.3} with $F_{WZ}=A_t=0.5$. We denote  the exclusion limits at 95\% C.L. in Refs.~\cite{Aaboud:2018cwk,Aad:2015typ,Khachatryan:2015qxa,
Sirunyan:2017sbn,Aad:2015nfa} of $\sigma(pp\to tH^\pm)\mathcal{B}(H^\pm\to tb)$ and $\sigma(pp\to jjH^\pm)\mathcal{B}(H^\pm\to W^\pm Z)$ as $[\sigma \mathcal{B}]_{tb}^{\text{limit}}$ and $[\sigma \mathcal{B}]_{WZ}^{\text{limit}}$, respectively. The upper limits can then  be re-interpreted as
\begin{align}
\label{eq:tb}
|A_t|^2/0.5^2\times \sigma(pp\to tH^\pm)_{\text{BM}}\mathcal{B}(H^\pm\to tb) &\leq  [\sigma \mathcal{B}]_{tb}^{\text{limit}},\\
\label{eq:WZ}
|F_{WZ}|^2/0.5^2\times \sigma(pp\to jjH^\pm)_{\text{BM}}\mathcal{B}(H^\pm\to W^\pm Z) &\leq [\sigma \mathcal{B}]_{WZ}^{\text{limit}},
\end{align}
where $\sigma(pp\to tH^\pm)_{\text{BM}}$ and $\sigma(pp\to jjH^\pm)_{\text{BM}}$ denote the LO cross sections~\footnote{The next-to-leading-order cross sections of charged Higgs boson production in this channel are available but model-dependent, see Refs.~\cite{Degrande:2015vpa,Kidonakis:2018qpw,Zaro:2015ika} and references therein.} of the $pp\to tH^\pm$ and the $pp\to jjH^\pm$, respectively. The constraints on $|F_{WZ}|^2\times \mathcal{B}(H^\pm\to W^\pm Z)$ and $|A_{t}|^2\times \mathcal{B}(H^\pm\to tb)$ are shown in Fig.~\ref{fig:constraints}. One can see that the constraints at the 13 TeV LHC are more stringent  than those at the 8 TeV LHC, except for $300\gev<m_{H^\pm}<450\gev$ in the process $pp\to jjH^\pm$, $H^\pm\to W^\pm Z$. In the following  analysis, we take the strongest constraints in the range of $m_{H^\pm}$ from 200~GeV to 1~TeV. On account of benchmark mass points in Ref.~\cite{Sirunyan:2017sbn}, we will choose the mass interval of 100~GeV to illustrate the combined sensitivities in Fig.~\ref{fig:couplings} and Fig.~\ref{fig:vev}.

With the minimal total width assumption, we can express the branching ratios $\mathcal{B}(H^\pm\to tb)$ and $\mathcal{B}(H^\pm\to W^\pm Z)$ in terms of the effective couplings $F_{WZ}$ and $A_t$. Therefore, we can obtain the discovery prospects and 95\% C.L. exclusion limits in the plane of $|F_{WZ}|$ and $|A_t|$. In Fig.~\ref{fig:couplings}, scenarios outside of the shaded regions are excluded at 95\% C.L. by the existing searches or our proposed search. The dashed and dot-dashed curves correspond to the discovery significance of $3\sigma$, $5\sigma$, respectively. For brevity, the processes $pp\to tH^\pm$, $H^\pm\to tb$ and $pp\to jjH^\pm$, $H^\pm \to W^\pm Z$ are denoted as $tb$ and $WZ$ modes, respectively. The process $pp\to jj H^\pm$, $H^\pm\to tb$ that depends on both the $H^\pm W^\mp Z$ and $H^\pm tb$ couplings is denoted as the mixed mode. We find that for $300\gev \lesssim m_{H^\pm}\lesssim 400\gev$ with $|F_{WZ}|,|A_t|\sim 0.5-1.0$, $H^\pm$ that couples to $W^\pm Z$ and $tb$ can be discovered in  the process $pp\to jj H^\pm$, $H^\pm\to tb$ at $\mathcal{Z}_D\geq 5$. For lighter or heavier charged Higgs boson, we can still achieve regions of the effective couplings that correspond to $\mathcal{Z}_D\geq 3$.  If $m_{H^\pm}\geq 500\gev$, top quark from the decay of $H^\pm$ is boosted. It is possible to improve the significance with jet substructure techniques~\cite{Aad:2014xra,Aad:2015typ}, which is beyond the scope of this study. On the other hand, if a null result is observed, 95\% C.L. exclusion limits in the plane of $|F_{WZ}|$ and $|A_t|$ are imposed. We obtain the most sensitive constraints on models with both $H^\pm W^\mp Z$ and $H^\pm tb$ couplings from the process $pp\to jj H^\pm$, $H^\pm\to tb$ for $|F_{WZ}|,|A_t|\sim 1.0$.

\section{Modified GM model with both \tf{$H^{\pm}W^{\mp}Z$}{HWZ} and \tf{$H^\pm tb$}{Htb} couplings}
\label{sec:model}

In this section, we will investigate a realistic model  with nonvanishing $H^{\pm}W^{\mp}Z$ and $H^\pm tb$ couplings. Recall that in the GM model~\cite{Georgi:1985nv,Chanowitz:1985ug}, there are two singly charged Higgs boson $H_3^{\pm}$ and $H_5^{\pm}$, which interact with fermions and $W^{\mp}Z$ separately since they transform in different custodial symmetry $SU(2)_C$ representations. It was found~\cite{Cen:2018wye,Haber:1999zh} that if  the custodial symmetry in the Higgs potential is relaxed, these two charged Higgs bosons can in general mix with each, resulting in two mass eigenstates that couple to both $W^\pm Z$ and $tb$. We will refer this model as the modified GM model. However, we emphasize that it is just a representative model. For example, in  a supersymmetric model~\cite{Bandyopadhyay:2015ifm,Bandyopadhyay:2017klv}, charged Higgs bosons can also couple to $W^{\pm}Z$ and fermions simultaneously due to the custodial symmetry breaking and doublet-triplet mixing.  Our results in Section~\ref{sec:effective} can be applied to any model with $H^{\pm}W^{\mp}Z$ and $H^\pm tb$ interactions. Below we will briefly introduce the modified GM model and show the implications in terms of the model parameters.

As in the original GM model~\cite{Georgi:1985nv,Chanowitz:1985ug}, we introduce two Higgs triplets: one real triplet $\xi\sim (3, 0)$ and one complex triplet $\chi\sim (3,1)$, which have the following forms:
\begin{eqnarray}
\chi=\left(
\begin{array}{cc}
\chi^{+}/\sqrt{2}&\chi^{++}\\
\chi^{0}&-\chi^{+}/\sqrt{2}
\end{array}\right)\;,\;\;\;\;\xi=\left(
\begin{array}{cc}
\xi^{0}/\sqrt{2}&\xi^{+}\\
\xi^{-}&-\xi^{0}/\sqrt{2}
\end{array}\right)\;
\end{eqnarray}
with the conventions $\xi^- = (\xi^+)^*$, $\chi^{0}=(v_{\chi}+h_{\chi}+iI_{\chi})/\sqrt{2}$ and $\xi^{0}=v_{\xi}+h_{\xi}$.

In order to discuss mass eigenstates of charged Higgs fields, it is convenient to use the basis $(G^+, H_3^+, H_5^+)$, 
\begin{align}
\left ( \begin{array}{c}
G^{+}\\
H_{3}^{+}\\
H_{5}^{+}
\end{array}
\right ) =
\left ( \begin{array}{ccc}
\dfrac{v_{H}}{v}&\dfrac{2v_{\xi}}{v}&-\dfrac{\sqrt{2}v_{\chi}}{v} \\[6pt]
-\dfrac{4v_{\xi}^{2}+2v_{\chi}^{2}}{N_{2}}&\dfrac{2v_{H}v_{\xi}}{N_{2}}&\dfrac{-\sqrt{2}v_{H}v_{\chi}}{N_{2}} \\[6pt]
0&\dfrac{\sqrt{2}v_{\chi}}{N_{3}}&\dfrac{2v_{\xi}}{N_{3}}
\end{array}
\right ) 
 \left ( \begin{array}{c}
h^+\\
\xi^+\\
\chi^+
\end{array}
\right )\;,
\end{align}
where $v\equiv \sqrt{v_H^2+4v_{\xi}^{2}+2v_{\chi}^{2}}\simeq 246\gev$ and the constants $N_i$ are given by $N_{2}=\sqrt{(4v_{\xi}^{2}+2v_{\chi}^{2})^{2}+4v_{H}^{2}v_{\xi}^{2}
+2v_{H}^{2}v_{\chi}^{2}}$ and $N_{3}=\sqrt{2v_{\chi}^{2}+4v_{\xi}^{2}}$. The Goldstone mode $G^+$ is ``eaten'' by $W^+$, while $H^+_3$ and $H^+_5$ are not mass eigenstates in general unless there is custodial symmetry in the Higgs potential. One needs to further diagonalize the mass matrix to obtain  the mass eigenstates $H^{m+}_3$ and $H^{m+}_5$,
\begin{eqnarray}
\left ( \begin{array}{c}
H^+_3\\
H^+_5
\end{array}
\right ) = 
\left (
\begin{array}{cc}
\cos \delta&\sin\delta\\
-\sin\delta&\cos\delta
\end{array}
\right )
\left ( \begin{array}{c}
H^{m+}_3\\H^{m+}_5
\end{array}
\right )\;, \label{charged-d}
\end{eqnarray}
The explicit form of the mixing angle $\delta$ can be found in Ref.~\cite{Cen:2018wye}. 

The interactions of  $H_3^{m\pm}$, $H_5^{m\pm}$ with $W^\mp Z$ and quarks are
\begin{eqnarray}
\label{eq:interactions}
&&\mathcal{L}_{W^\pm Z} = ({g^2\over 2 c_W} {v_H(2v^2_\chi - 4 v^2_\xi)\over N_2} \cos\delta  + {g^2\over 2c_W} {4\sqrt{2}  v_\chi v_\xi\over {N_3}} \sin\delta )H_3^{m+} W^-_\mu Z^\mu\nonumber\\
&&\;\;\;\;\;\;\;\;\;\;\;+({g^2\over 2 c_W} {v_H(2v^2_\chi - 4 v^2_\xi)\over N_2} \sin\delta  - {g^2\over 2c_W} {4\sqrt{2}  v_\chi v_\xi\over {N_3}} \cos\delta )H_5^{m+} W^-_\mu Z^\mu+h.c.\;,\\
&&\mathcal{L}^q_{\text{Yuk}}
=-\sqrt{2}\frac{1}{v_H}\frac{4v_\xi^2+2v_\chi^2}{N_2}  (\bar{U} \hat  M_u V_\text{{CKM}} P_L D- \bar{U}V_\text{{CKM}} \hat M_d  P_R D)\nonumber\\
&&\;\;\;\;\;\;\;\;\;\;\;\times ( \cos\delta H_3^{m+}+ \sin\delta H_5^{m+}) +h.c.\;,\nonumber
\end{eqnarray}
where $c_W\equiv \cos\theta_W$, $U = (u, c, t)^T$, $D = (d, s, b)^T$ and $V_\text{{CKM}}$ denotes the Cabibbo-Kobayashi-Maskawa matrix. The $\rho$ parameter is expressed as
\begin{eqnarray}
\label{eq:rho}
\rho = {v^2_H + 2 v^2_\chi + 4 v^2_\xi\over v^2_H + 4 v^2_\chi}\;. 
\end{eqnarray}
With custodial symmetry, $v_\xi = v_\chi/\sqrt{2}$ and $\sin\delta=0$, thus $\rho=1$ and $H^{m\pm}_5(=H_{5}^{\pm})$ does not couple to quarks, whereas $H^{m\pm}_3(=H_{3}^{\pm})$ does not couple to  $W^\mp Z$ at tree level.

Since the $\rho$ parameter is severely constrained, $\rho^{\text{exp}} = 1.00039\pm 0.00017$~\cite{Tanabashi:2018oca},   the $v_\chi,v_\xi$ are generally constrained to be less  than a few GeV if they are independent. However when $v_\xi = v_\chi/\sqrt{2}$,  one has $\rho=1$, and larger $v_\chi$ and $v_\xi$ are allowed. In this case, $\tan2\delta$ is proportional to the combination $2v_{\chi}(-\sqrt{2}\mu_{\chi HH}+\mu_{\xi HH})+v_{\chi}^2 (\kappa_3+\sqrt{2}\lambda)$, where $\mu_{\chi HH}$, $\mu_{\xi HH}$, $\kappa_3$ and $\lambda$ are the couplings in the general Higgs potential~\cite{Cen:2018wye}~\footnote{If  the custodial symmetry  is removed, these couplings are independent, while in the GM model, $\mu_{\xi HH}=\sqrt{2}\mu_{\chi HH}$ and $\kappa_3=-\sqrt{2}\lambda_5$~\cite{Cen:2018wye}.}. In the GM model this combination is forced to be zero by the custodial symmetry. However, if the custodial symmetry is broken explicitly the mixing angle $\delta$ can be sizable.  We will work in the modified GM (MGM) model with the working hypothesis $v_\xi = v_\chi/\sqrt{2}$. 

\begin{figure}[!htb]
\centering
\includegraphics[width=0.3\textwidth]{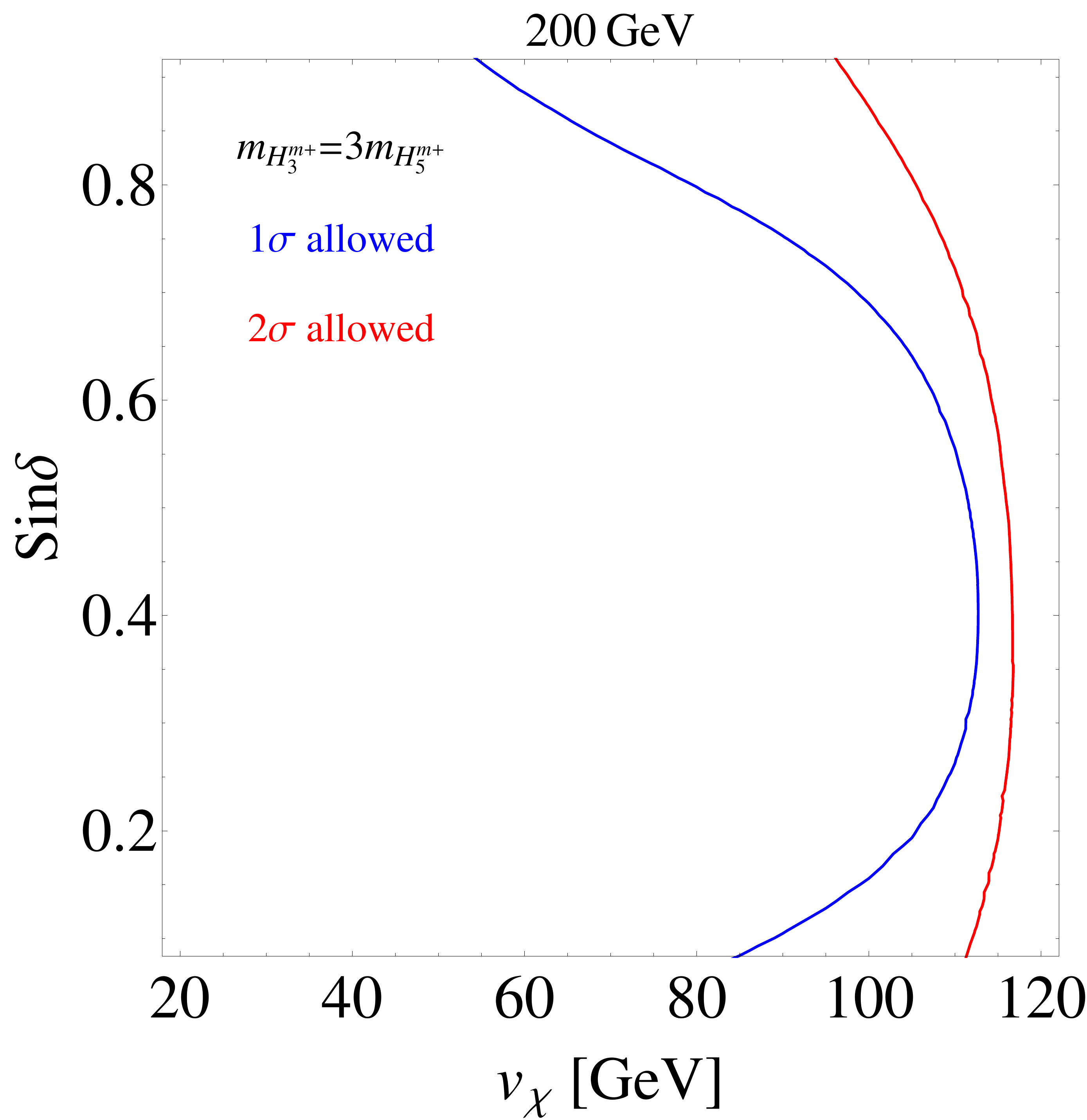}	
\includegraphics[width=0.3\textwidth]{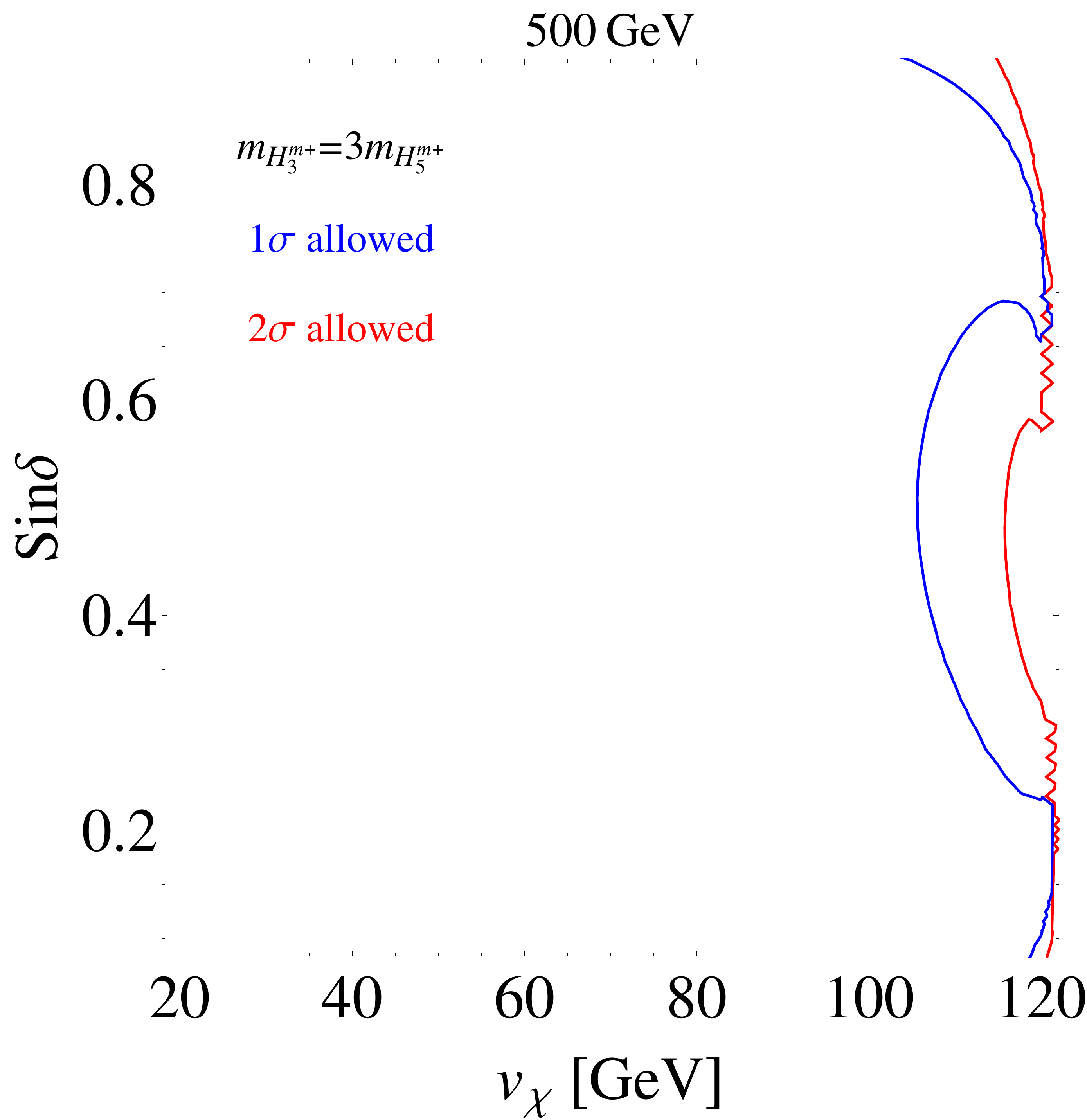}	
\includegraphics[width=0.3\textwidth]{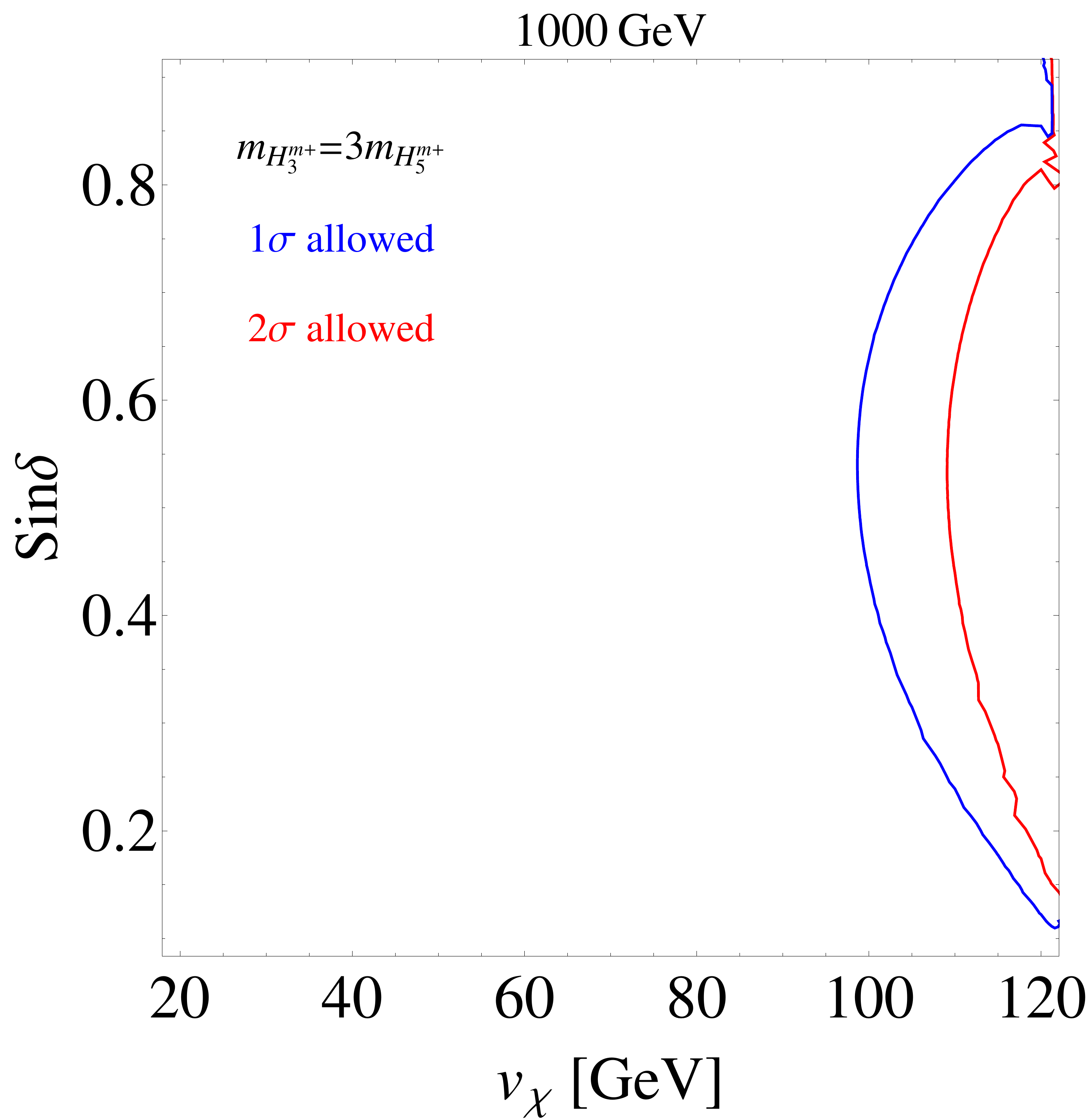}		
\caption{Constraints on the triplet VEV $v_{\chi}$ as well as sine of the mixing angle $\delta$ in the MGM model with the assumption $m_{H^{m+}_{3}}=3m_{H^{m+}_{5}}$. The mass of charged Higgs boson $H_5^{m+}$ is chosen to be $200,500,1000\gev$ from left to right panels. The region on the left of the blue (red) curve is allowed at $1\sigma$~$(2\sigma)$ level.}
\label{fig:Zbb}
\end{figure}

The interactions of charged Higgs bosons to quarks and $W^{\pm}Z$ in the MGM model can be expressed as
\begin{align}
&\mathcal{\mathcal{L}}_{W^\pm Z} =\dfrac{g^2v_{\chi}}{c_W}W^-_\mu Z^\mu   (\sin\delta  H_3^{m+}-\cos\delta  H_5^{m+} )+h.c.,\\
&\mathcal{L}^q_{\text{Yuk}} =-\dfrac{2\sqrt{2}v_{\chi}}{v_H v}(\bar{U} \hat  M_u V_{\text{CKM}} P_L D- \bar{U}V_\text{{CKM}} \hat M_d  P_R D) (\cos\delta  H_3^{m+}+\sin\delta  H_5^{m+} )+h.c.,
\end{align}
where 
\begin{align}
N_2=2v_{\chi}v,\quad N_3=2v_{\chi},\quad v=\sqrt{v_H^2+4v_{\chi}^2}.
\end{align}
The mixing angle $\delta$ is in the range of $[0,2\pi)$, so $H_3^{m\pm}$ and $H_5^{m\pm}$ can interact with both quarks and $W^\pm Z$ unless $\delta=n\pi/2$ with $n=0,1,2,3$.

Compared with the Lagrangian in Eq.~\eqref{eq:couplings}, we can obtain the effective couplings $F_{WZ}$, $A_t$ and $A_b$ as follows:
\begin{align}
\label{eq:coup-H3}
F_{WZ}=\dfrac{gv_{\chi}}{c_W m_W}\sin\delta,\quad A_t=-A_b=\dfrac{2v_{\chi}}{v_{H}}V_{tb}\cos\delta,
\end{align}
for the  $H_3^{m\pm}$ and 
\begin{align}
\label{eq:coup-H5}
F_{WZ}=-\dfrac{gv_{\chi}}{c_W m_W}\cos\delta,\quad A_t=-A_b=\dfrac{2v_{\chi}}{v_{H}}V_{tb}\sin\delta,
\end{align}
for the $H_5^{m\pm}$. Combined with the effective Lagrangian in Eq.~\eqref{eq:couplings}, it is apparent that the magnitudes of the right-handed $H^\pm tb$ couplings are suppressed and smaller than the left-handed one by a factor of $m_b/m_t$.

The couplings of $H_3^{m^\pm}$ and $H_5^{m^\pm}$ to quarks and $W^\mp Z$ are proportional to the triplet VEV $v_{\chi}$, which is a generic feature of the GM-type models~\cite{Logan:2015xpa}.  A larger triplet VEV can result in a better sensitivity. With the sum rule $v_H^2+4v_{\chi}^2=v^2$ in the MGM model, the triplet VEV $v_{\chi}=\sqrt{v^2-v_{H}^2}/2\lesssim 88,121,123\gev$ is required in order to satisfy the perturbative bound of the top Yukawa coupling $m_t/v_H \lesssim 1,\sqrt{4\pi},4\pi$ from the renormalization group running~\cite{Machacek:1983fi,Altmannshofer:2013zba}. Moreover, in general we can obtain a strict upper bound on the triplet VEV from the perturbative unitarity of the scattering amplitude for $t\bar{t}\to t\bar{t}$~\cite{Chanowitz:1978uj,Murdock:2008rx}. One can obtain that $y_t\lesssim \sqrt{16\pi/5}$, which implies that the doublet VEV $v_{H}\gtrsim 78\gev$. In the MGM model, the couplings of the neutral Higgs bosons to $t\bar{t}$ also depend on other parameters compared with the model discussed in Refs.~\cite{Chanowitz:1978uj,Murdock:2008rx}. But the constrain can still provide some guide for the allowed Yukawa coupling, so that the triplet VEV $v_{\chi}\lesssim 117\gev$.

\begin{figure}[!htb]
\centering
\includegraphics[width=0.3\textwidth]{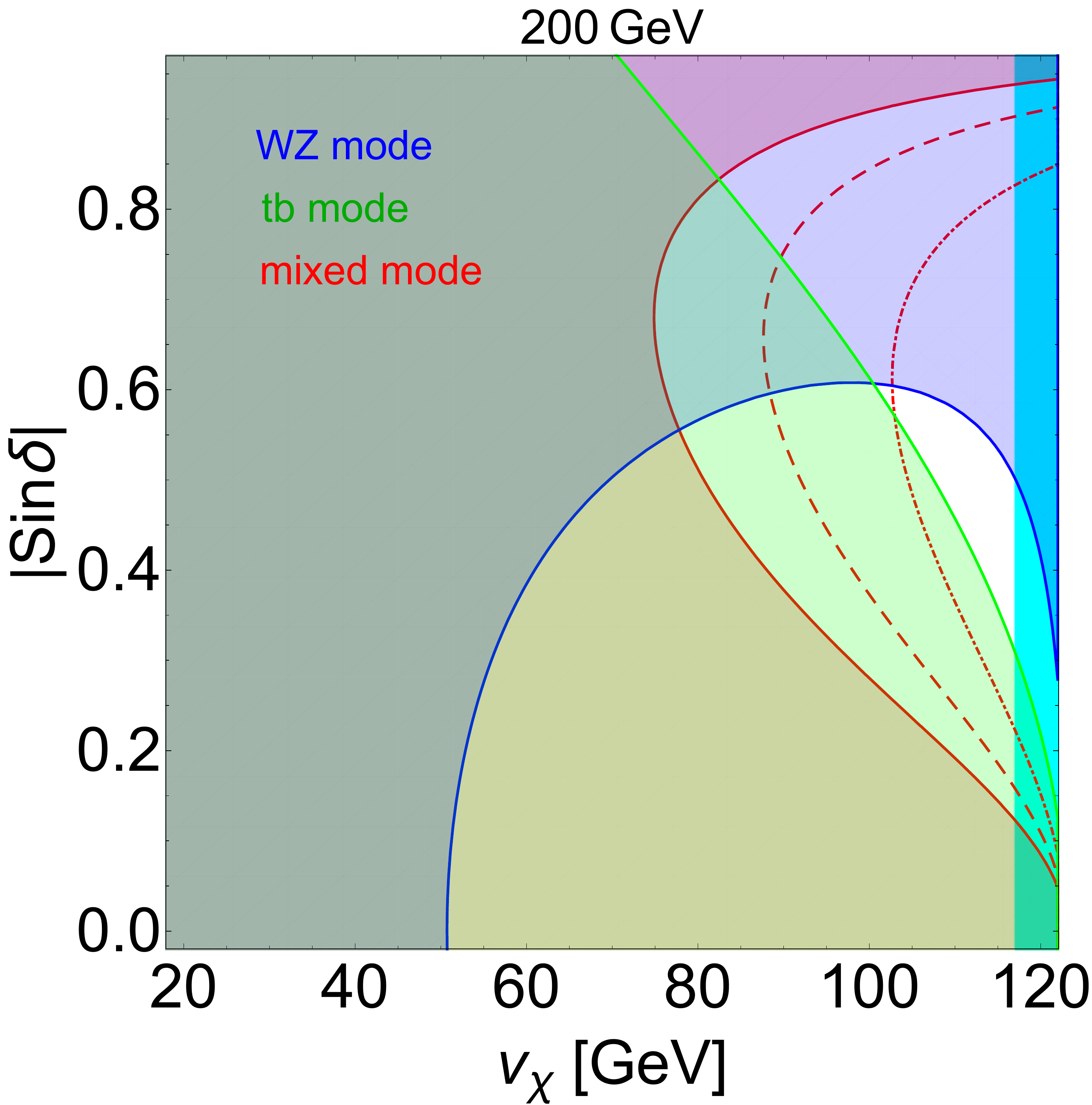}	
\includegraphics[width=0.3\textwidth]{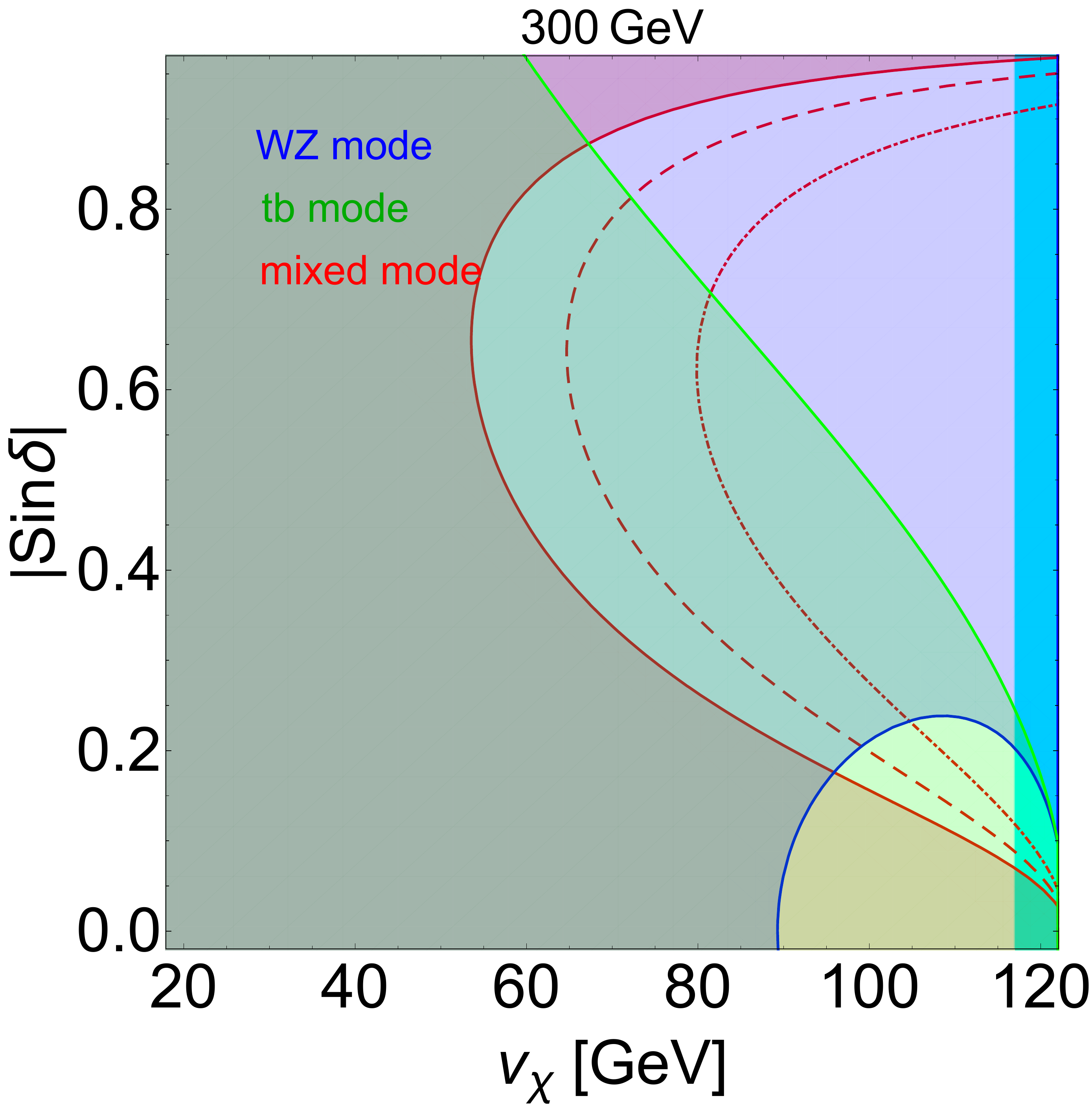}	
\includegraphics[width=0.3\textwidth]{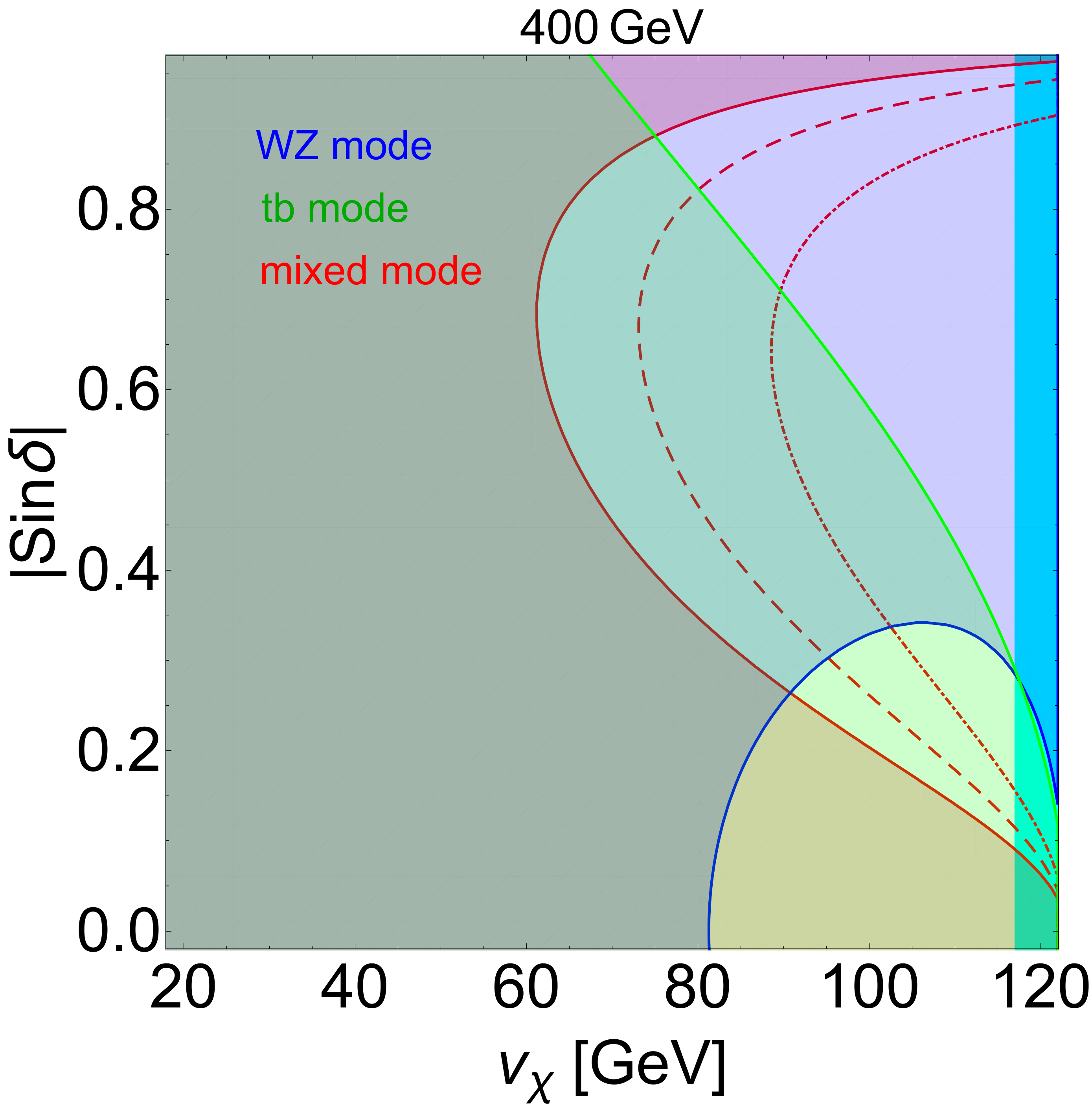}	
\includegraphics[width=0.3\textwidth]{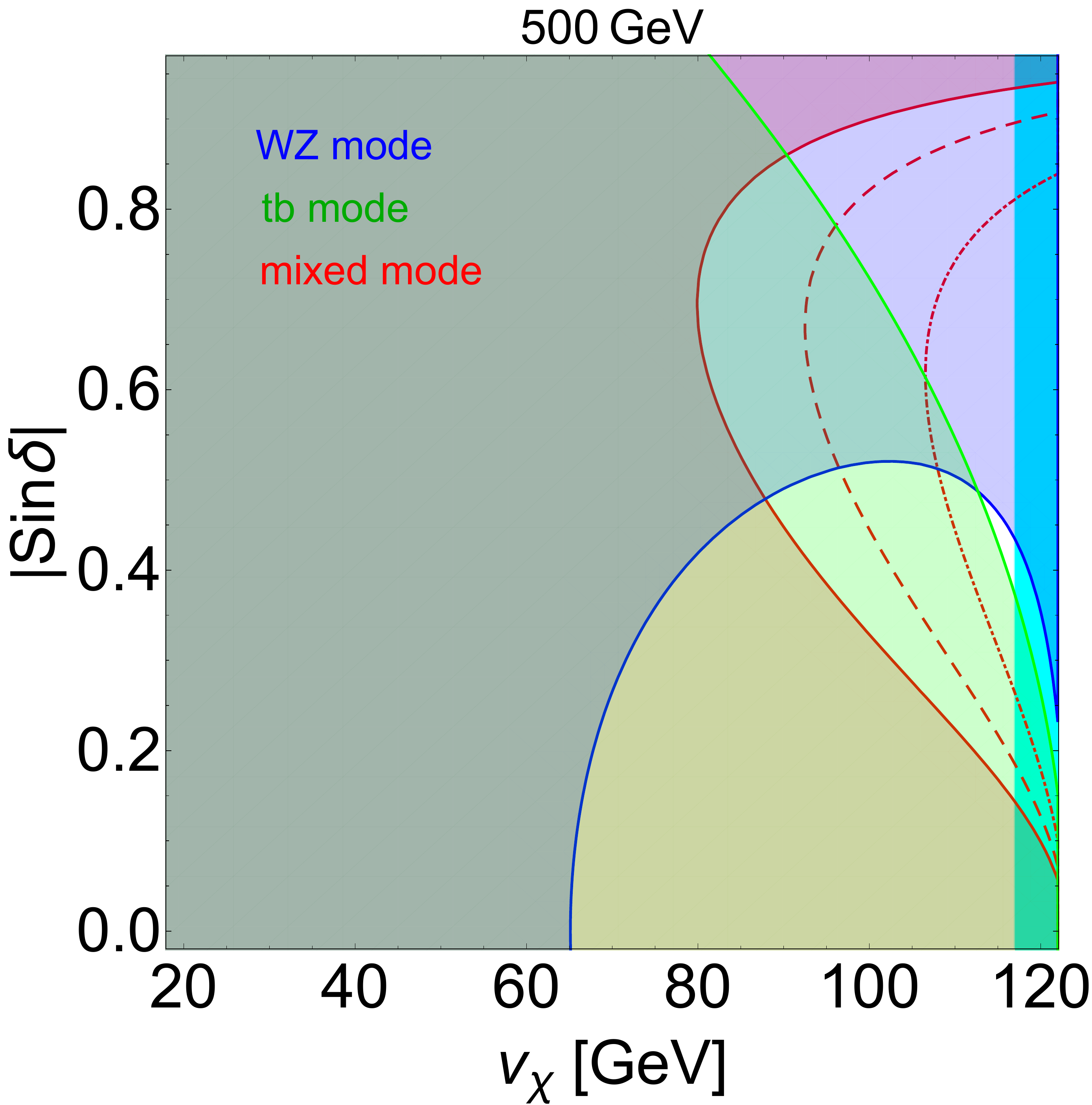}	
\includegraphics[width=0.3\textwidth]{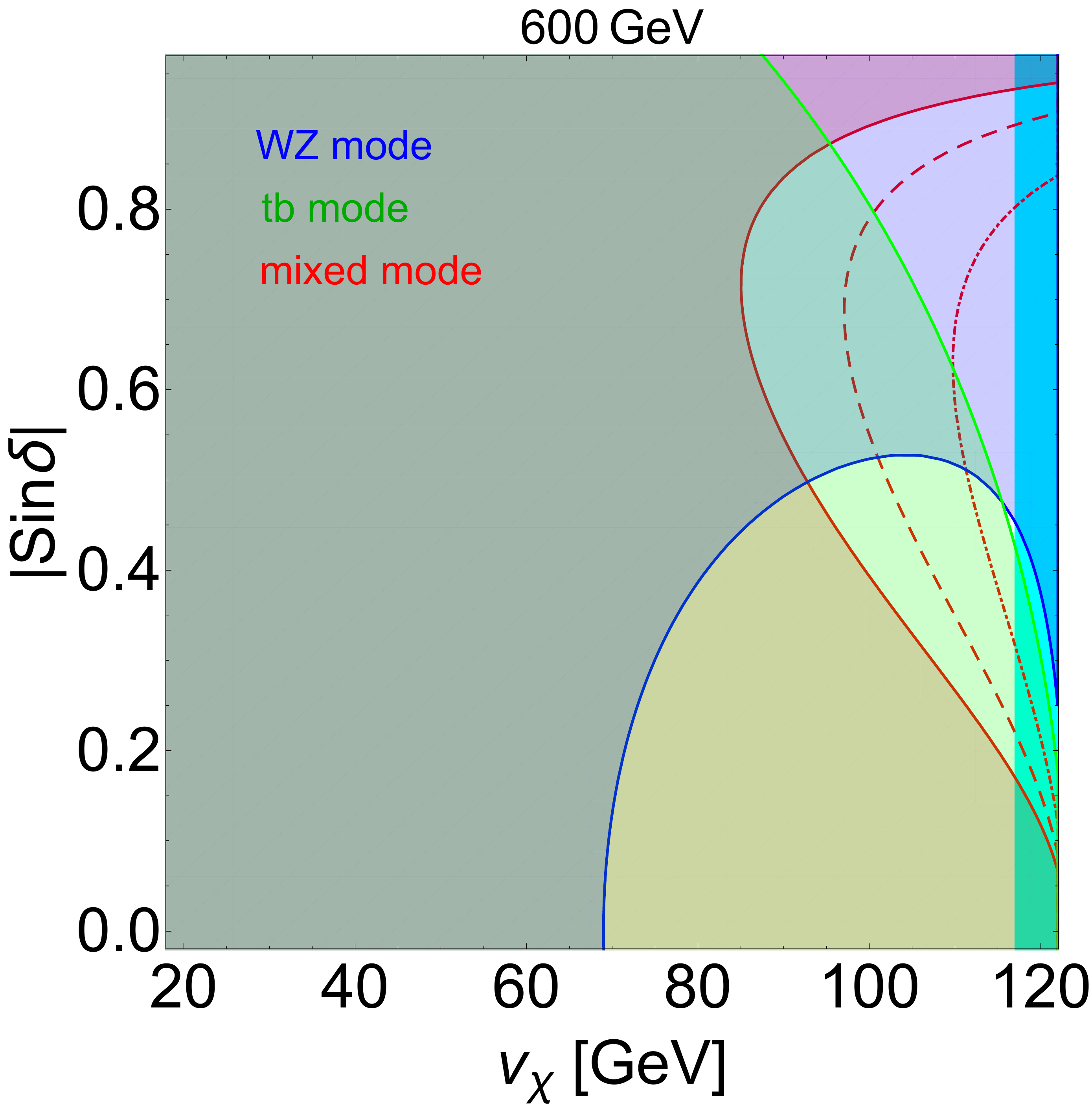}	
\includegraphics[width=0.3\textwidth]{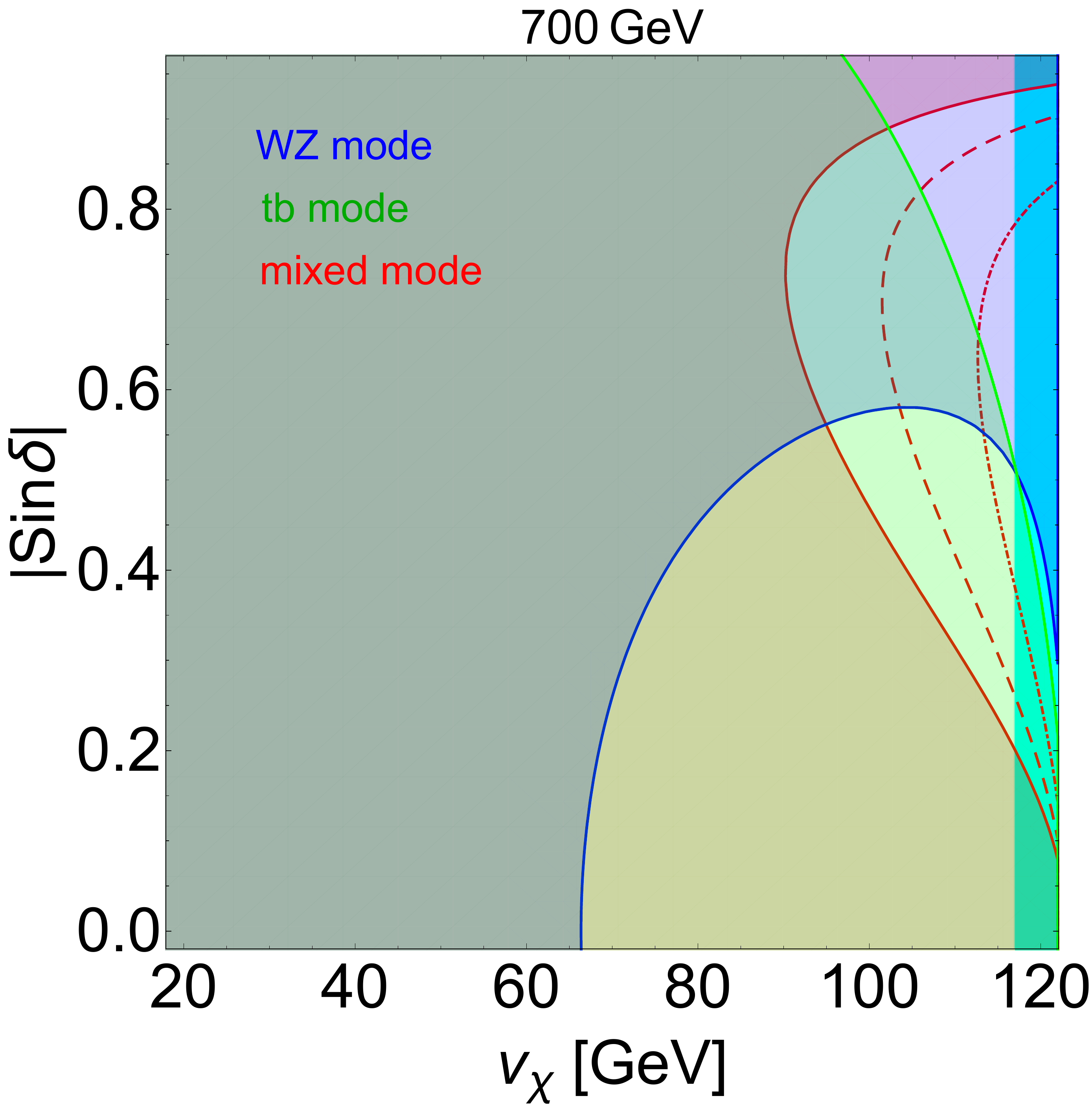}	
\includegraphics[width=0.3\textwidth]{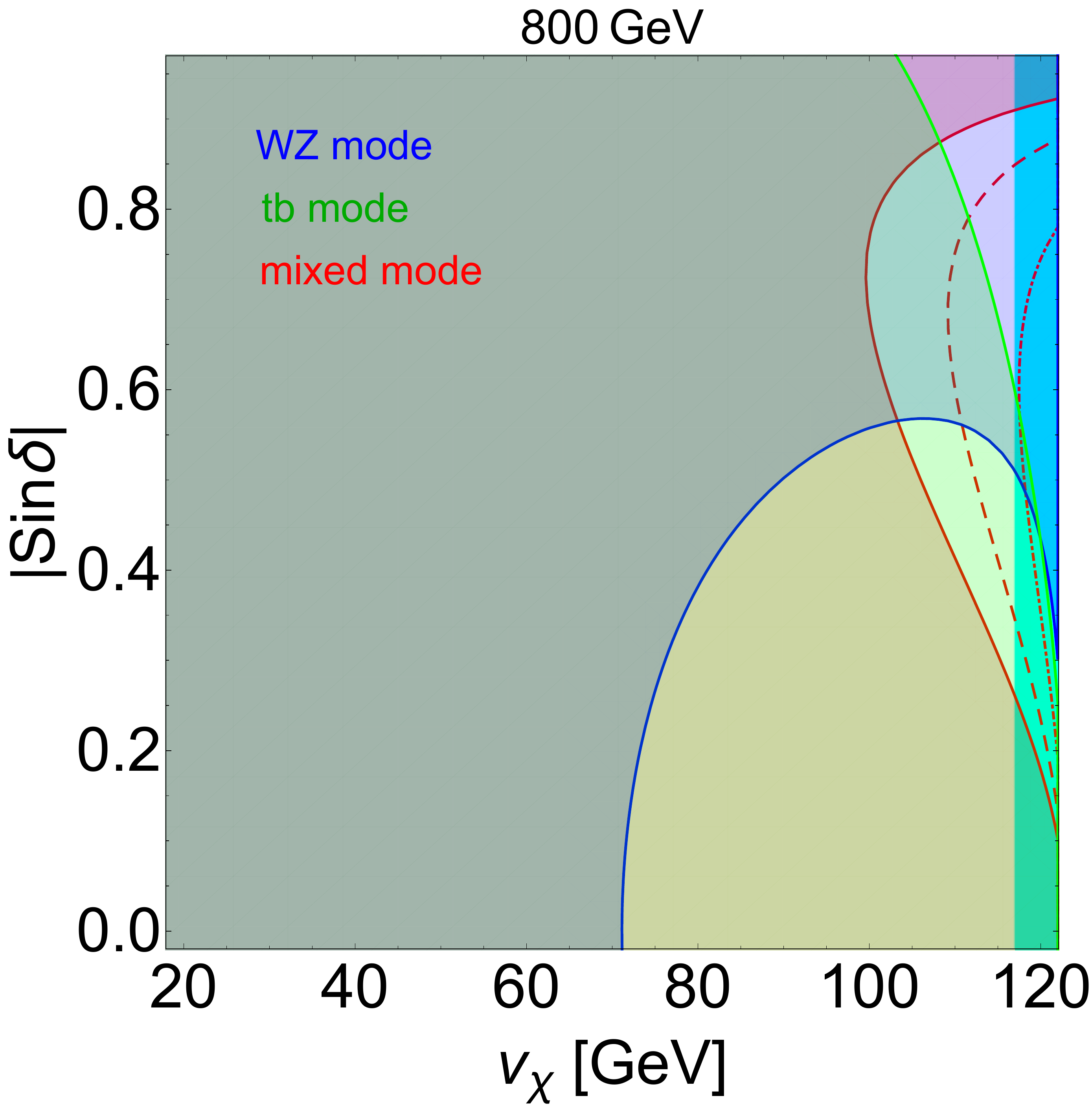}	
\includegraphics[width=0.3\textwidth]{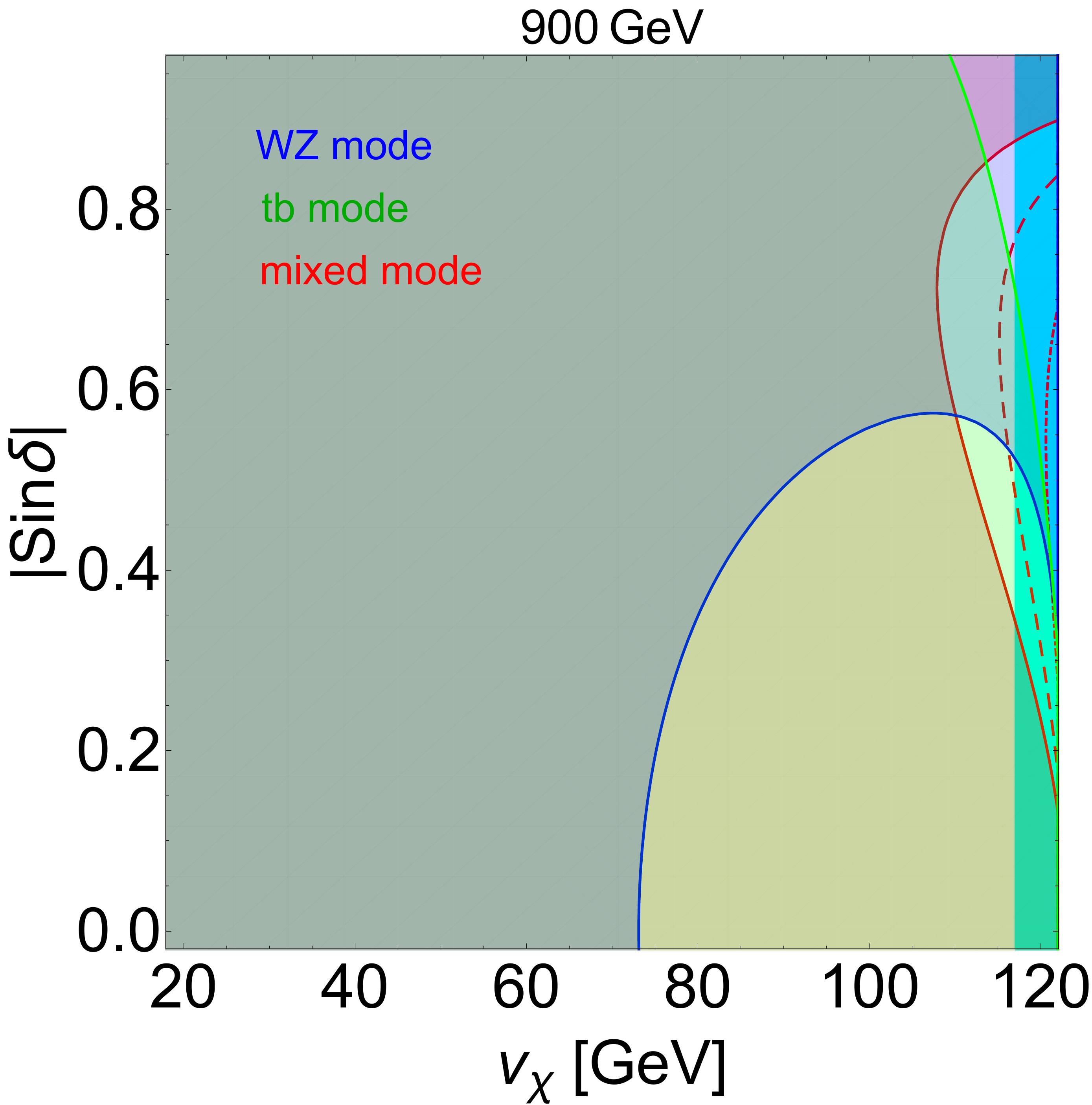}	
\includegraphics[width=0.3\textwidth]{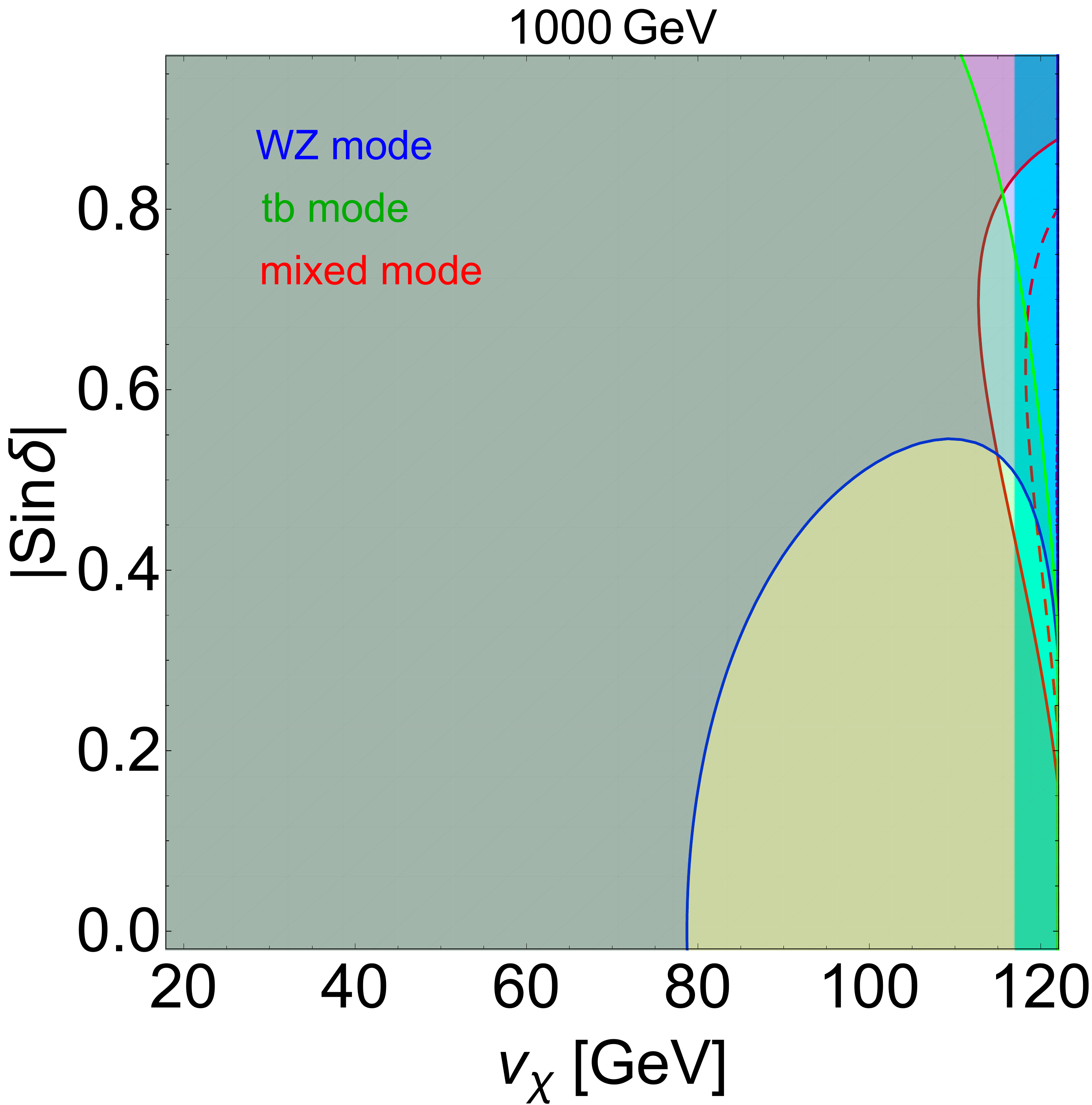}	
\caption{Sensitivities to $v_{\chi}$ and $|\sin\delta|$ in the MGM model with $H^\pm=H_5^{m\pm} $ for $m_{H_5^{m^\pm }}=200-1000\gev$ with the minimal total width assumption. The blue and green regions are obtained from the exclusion limit at 95\% C.L. of the processes $pp\to tH^\pm$, $H^\pm\to tb$ and $pp\to jjH^\pm$, $H^\pm \to W^\pm Z$, respectively. The red solid, dashed and dot-dashed curves correspond to the exclusion limit at 95\% C.L. and discovery prospects with $\mathcal{Z}_D=3,5$, respectively. The cyan vertical regions on the right of the panels are excluded by the perturbative unitarity requirement.}
\label{fig:vev}
\end{figure}

On the other hand, the $H_3^{m\pm }tb$ and $H_5^{m\pm} tb$ couplings can also modify the $Zb\bar{b}$ coupling through higher order corrections. At the 1-loop level, the correction to the left-handed coupling is expressed as~\cite{Haber:1999zh,Logan:1999if} 
\begin{align}
\delta g_{\text{MGM}}^{L}&=\dfrac{1}{32\pi^2}\dfrac{e}{s_Wc_W}(\dfrac{gm_t}{\sqrt{2}m_W})^2\tan^2\theta_H \bigg \{ \sin^2\delta \big[\dfrac{R_5}{R_5-1}-\dfrac{R_5\log R_5}{(R_5-1)^2}\big]\nn\\
& +\cos^2\delta \big[\dfrac{R_3}{R_3-1}-\dfrac{R_3\log R_3}{(R_3-1)^2}\big]\bigg\}
+\dfrac{1}{16\pi^2}\dfrac{e}{s_Wc_W}(\dfrac{gm_t}{\sqrt{2}m_W})^2\tan^2\theta_H \nn\\
&\times(2\cos\theta_H \sin\delta\cos\delta ) \bigg\{C_{24}(m_t^2,m_W^2,m_{H_3^{m+}}^2)-C_{24}(m_t^2,m_W^2,m_{H_5^{m+}}^2)\nn\\
&+\sin^2\delta [C_{24}(m_t^2,m_{H_5^{m+}}^2,m_{H_5^{m+}}^2)-C_{24}(m_t^2,m_{H_5^{m+}}^2,m_{H_3^{m+}}^2)]\nn\\
&+\cos^2\delta [C_{24}(m_t^2,m_{H_5^{m+}}^2,m_{H_3^{m+}}^2)-C_{24}(m_t^2,m_{H_3^{m+}}^2,m_{H_3^{m+}}^2)]
\bigg\},
\end{align}
where $R_3=m_t^2/m_{H_3^{m+}}^2$, $R_5=m_t^2/m_{H_5^{m+}}^2$, $\tan\theta_H\equiv{2v_{\chi}}/{v_H}$ and $\cos\theta_H\equiv v_H/v$.  The first three arguments $(m_b^2,m_Z^2, m_b^2)$ of the three-point integrals $C_{24}$~\cite{Logan:1999if} are dropped for simplicity. The first term is positive definite while the second term can be positive or negative. If $\sin\delta\cos\delta>0$, the second term is negative for $m_{H_3^{m+}}>m_{H_5^{m+}}$ and cancels the contribution from the first term. On the other hand, the correction to the right-handed coupling $\delta g_{\text{MGM}}^{R}$ is proportional to $b$-quark mass and negligible. The modifications of the $Zb\bar{b}$ coupling have  been measured in the hadronic branching ratio $R_b\equiv \Gamma(Z\to b\bar{b})/\Gamma(Z\to \text{hadronic})$~\cite{Tanabashi:2018oca},
\begin{align}
\label{eq:Rb_exp}
R_{b}^{\text{exp}}=0.21629\pm 0.00066.
\end{align}
The theoretical $R_b$ in the MGM model is expressed as
\begin{align}
R_b^{\text{MGM}}=R_{b}^{\text{SM}}+\delta R_b^{\text{MGM}},
\end{align}
where the SM value $R_{b}^{\text{SM}}=0.2158$~\cite{Tanabashi:2018oca}, $\delta R_b^{\text{MGM}}$ that depends on $\delta g_{\text{MGM}}^{L}$ can be parameterized as~\cite{Haber:1999zh}
\begin{align}
\delta R_b^{\text{MGM}}=-0.7785\delta g_{\text{MGM}}^{L}.
\end{align}
Again, $\delta g_{\text{MGM}}^{R}$ is neglected.
Given the values of $m_{H_5^{m+}}$ and $m_{H_5^{m+}}$, $R_{b}^{\text{MGM}}$ only depends on $v_{\chi}$ and $\sin\delta$. The $1\sigma$  and $2\sigma$ constraints from $R_b$ measurements are implemented by requiring 
\begin{align}
|R_{b}^{\text{MGM}}-(R_{b}^{\text{exp}})_{\text{cent}}|<(R_{b}^{\text{exp}})_{\text{err}},2\times(R_{b}^{\text{exp}})_{\text{err}}.
\end{align} 
Here the subscripts ``cent'' and ``err'' denote the central value and error of $R_{b}^{\text{exp}}$ in Eq.~\eqref{eq:Rb_exp}. 

\begin{figure}[!htb]
\centering
\includegraphics[width=0.4\textwidth]{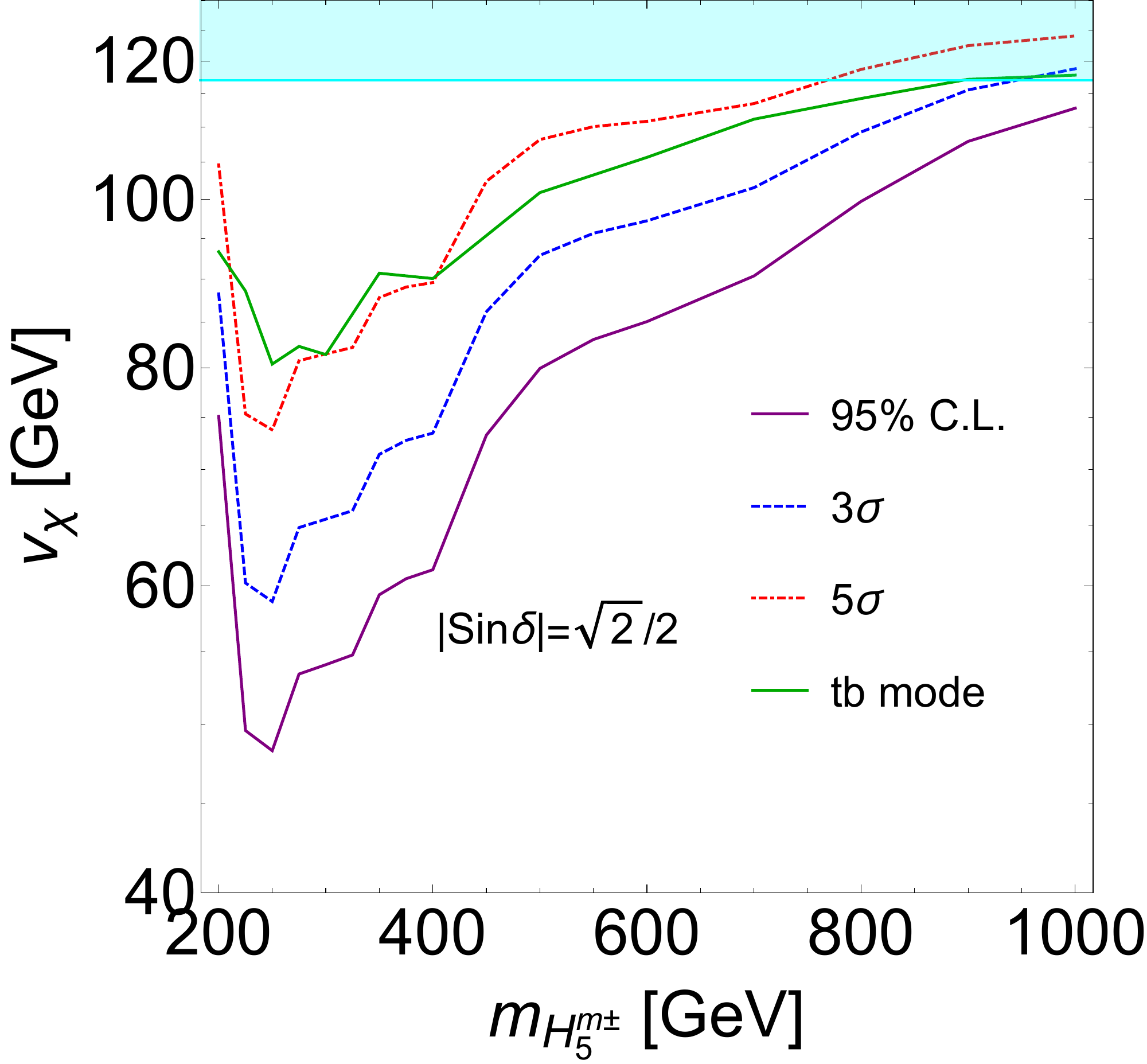}		
\caption{Sensitivities to the $m_{H^{m\pm}_{5}}$ and $v_{\chi}$ in the MGM model in the mass range from 200~GeV to 1~TeV with the assumption $|\sin\delta|=1/\sqrt{2}$.  The cyan horizontal region is excluded by the perturbative unitarity requirement.}
\label{fig:mhp-vev}
\end{figure}

In Fig.~\ref{fig:Zbb}, we show the $1\sigma,2\sigma$ contours in the plane of $v_{\chi}$ and $\sin\delta$ for $m_{H_5^{m+}}=200$, $500$, $1000\gev$ with the assumption $m_{H_3^{m+}}=3m_{H_5^{m+}}$. 
 We find  that $v_{\chi}\lesssim 100\gev$ is still allowed by the perturbative bound and the $Zb\bar{b}$ data, while the exact upper limit depends on the mixing angle $\delta$ and the interplay of two charged Higgs bosons $H_3^{m\pm}$ and $H_5^{m\pm}$. Therefore, we do not show the $Zb\bar{b}$ constraint explicitly in the sensitivity plots in Fig.~\ref{fig:vev} and Fig.~\ref{fig:mhp-vev}.

In Section~\ref{sec:effective}, we have obtained the model-independent sensitivities to the effective $H^\pm W^\mp Z$ and $H^\pm tb$ couplings. Having the explicit forms of $F_{WZ}$ and $A_t$ in the MGM model in Eqs.~\eqref{eq:coup-H3}~\eqref{eq:coup-H5}, we can obtain the sensitivities to the model parameters. i.e., the mixing parameter $\delta$ and the triplet VEV $v_{\chi}$. We assume that the charged Higgs boson $H^\pm$ is identified as $H^{m\pm}_{5}$, and it mainly decays into $W^\pm Z$ and $tb$. In the mass range from 200~GeV to 1~TeV, sensitivities to $v_{\chi}$ and $|\sin\delta|$ are shown in Fig.~\ref{fig:vev}. In accord with   Fig.~\ref{fig:couplings}, the charged Higgs boson $H_5^{m^\pm}$ can be discovered $(\mathcal{Z}_D\geq 5)$ in the process $pp\to jjH^\pm$, $H^\pm \to tb$ for $300\gev \lesssim m_{H_5^{m\pm}}\lesssim 400\gev$ with the triplet VEV $80\gev\lesssim v_{\chi}\lesssim 100\gev$. On the other hand, if the $H_5^{m^\pm}$ is not found, one can put 95\% C.L. exclusion limits of the signal process in the plane of $v_{\chi}$ and $|\sin\delta|$. 
Especially, $v_{\chi}\lesssim 53,62\gev$ for $|\sin\delta|\simeq 0.7$ can be achieved with $m_{H_5^{m\pm}}=300,400\gev$, which are well below the upper limits $90,80\gev$ for $|\sin\delta|=0$, respectively. To evaluate the sensitivities to the triplet VEV as a function of the charged Higgs boson mass. We choose that $|\sin\delta|=1/\sqrt{2}$, and obtain the sensitivities in the plane of $m_{H^{m\pm}_5}$ and $v_{\chi}$ in Fig.~\ref{fig:mhp-vev}. The current constraint obtained from the exclusion limit in the process $pp\to tH^\pm$, $H^\pm\to tb$ is shown in the green curve, while there is no useful constraint from the process $pp\to jjH^\pm$, $H^\pm\to W^\pm Z$. The $3\sigma,5\sigma$ discovery prospects and exclusion limit at 95\% C.L. in the process $pp\to jjH^\pm$, $H^\pm \to tb$ are also shown. Both of the processes $pp\to jjH^\pm$, $H^\pm \to tb$ and $pp\to tH^\pm$, $H^\pm\to tb$ get the best sensitivities for $m_{H_5^{m\pm}}\simeq 250\gev$ resulting dips in the curves. We obtain that the sensitivity to the triplet VEV $v_{\chi}$ of our proposed search is better than that of the existing searches. For the  $m_{H^{m\pm}_5}\lesssim 500\gev$, the region of the triplet VEV $ v_{\chi}\lesssim 80\gev$  can be probed at 95\% C.L., shown  as the purple curve in Fig.~\ref{fig:mhp-vev}.

Before enclosing this section, we emphasize that the MGM model is only a representative  model. Other models with triplets or higher $SU(2)_L$ representation of Higgs multiplets, in which charged Higgs boson(s) can couple to fermions and $W^\pm Z$ simultaneously,   can be studied similarly.

\section{Conclusions}
\label{sec:summary}

In this work, we  have extended  the existing searches of charged Higgs boson at the LHC. Different from the processes inspired by the 2HDMs and GM model, the VBF process $pp\to jjH^\pm$, $H^\pm\to tb$ requires the existence of both $H^\pm W^\mp Z$ and $H^\pm tb$ couplings. We have performed a model-independent analysis of this process at the LHC with the effective $H^\pm W^\mp Z$ and $H^\pm tb$ couplings.  With the minimal total width assumption, we interpret the results in terms of the effective couplings $F_{WZ}$ and $A_t$ for $m_{H^\pm}$ in the range from 200~GeV to 1~TeV.  We found that the process $pp\to jj H^\pm$, $H^\pm\to tb$ can be  discovered $(\mathcal{Z}_D\geq 5)$ for $300\gev \lesssim m_{H^\pm}\lesssim 400\gev$ with $|F_{WZ}|,|A_t|\sim 0.5-1.0$. Discovering  the process in the region of $m_{H^\pm}\geq 500\gev$ requires the improved experimental selection criteria. However, one can still obtain the most sensitive constraints on models with both $H^\pm W^\mp Z$ and $H^\pm tb$ couplings through this process.

We investigate the implications in a realistic model, the MGM model, which introduces two Higgs triplets into the SM analogous to the GM model. Since the requirement of custodial symmetry in the Higgs potential after the EWSB is relaxed, two physical singly charged Higgs bosons $H_3^{m^\pm}$ and $H_5^{m^\pm}$ with both couplings to quarks and $W^\pm Z$ are achieved. We discuss the theoretical as well as experimental constraints of this model. Then the sensitivities to the model parameters, i.e., the triplet VEV $v_{\chi}$ and the mixing angle $\delta$ are obtained and compared with constraints from the existing searches if the charged Higgs boson $H^\pm$ is identified as $H_5^{m\pm}$. We have pointed out  that $H_5^{m^\pm}$ can be discovered in the process $pp\to jjH^\pm$, $H^\pm \to tb$ for $300\gev \lesssim m_{H_5^{m\pm}}\lesssim 400\gev$ with $80\gev\lesssim v_{\chi}\lesssim 100\gev$. Supposing a maximal mixing pattern with $|\sin\delta|= 1/\sqrt{2}$, the exclusion limit at 95\% C.L. on the triplet VEV $v_{\chi}\lesssim 80\gev$ can be achieved for $m_{H_5^{m\pm}}\lesssim 500\gev$.

Finally, the signal process proposed in this work is a direct evidence for a charged Higgs boson that couples to fermions and $W^\pm Z$, which is complementary to current searches for charged Higgs bosons.
Our study in this work can be used as a roadmap of future charged Higgs boson searches at the LHC.

\begin{acknowledgments}
We would like to thank Ying-nan Mao, Minho Son, Yi-Lei Tang, Ke-Pan Xie, Kei Yagyu and Bin Yan for helpful discussions. 
This work was supported in part by the MOST (Grant No. MOST 106-2112-M-002-003-MY3 ), and in part by Key Laboratory for Particle Physics,
Astrophysics and Cosmology, Ministry of Education, and Shanghai Key Laboratory for Particle
Physics and Cosmology (Grant No. 15DZ2272100), and in part by the NSFC (Grant Nos. 11575110, 11575111, 11655002,  and 11735010).
\end{acknowledgments}

%%%%%%%%%%%%%%%%%%%%%%%%%%%%%%%%%%%%%%%
%%%%%%%%%%%%%%%%%%%%%%%%%%%%%%%%%%%%%%%

%\bibliographystyle{apsrev}
\bibliographystyle{JHEP}
\bibliography{reference}
%
%%%%%%%%%%%%%%%%%%%%%%%%%%%%%%%%%%%%%%%%%%%%%
%%%%%%%%%%%%%%%%%%%%%%%%%%%%%%%%%%%%%%%%%%%%%
%%%%%%%%%%%%%%%%%%%%%%%%%%%%%%%%%%%%%%%%%%%%%%5
\end{document}